\documentclass[aps,pra,reprint,superscriptaddress,showpacs,longbibliography]{revtex4-1}
\usepackage{amssymb,graphicx,epsfig,threeparttable}
\usepackage{color}
\usepackage[colorlinks=true,linkcolor=blue]{hyperref}
\usepackage{ulem}

\begin{document}

\title{Vibration-induced coherence enhancement of the performance of a biological quantum heat engine}
\author{Hong-Bin Chen}
\affiliation{Department of Physics, National Cheng Kung University, Tainan 701, Taiwan}
\author{Pin-Yi Chiu}
\affiliation{Department of Physics, National Cheng Kung University, Tainan 701, Taiwan}
\author{Yueh-Nan Chen}
\email{yuehnan@mail.ncku.edu.tw}
\affiliation{Department of Physics, National Cheng Kung University, Tainan 701, Taiwan}
\affiliation{Physics Division, National Center for Theoretical Sciences, Hsinchu 300, Taiwan}

\date{\today}

\begin{abstract}
Photosynthesis has been a long-standing research interest due to its fundamental importance. Recently, studies on photosynthesis processes also have inspired attention
from a thermodynamical aspect when considering photosynthetic apparatuses as biological quantum heat engines. Quantum coherence is shown to play a crucial role in enhancing
the performance of these quantum heat engines. Based on the experimentally reported structure, we propose a quantum heat engine model with a non-Markovian vibrational mode.
We show that one can obtain a performance enhancement easily for a wide range of parameters
in the presence of the vibrational mode. Our results provide insights into the photosynthetic processes and a design principle mimicking natural organisms.
\end{abstract}

\maketitle

\section{INTRODUCTION}

Photosynthesis, which occurs naturally in green plants, bacteria, and algae, harvests solar energy and converts it into chemical energy with approximately 100\% quantum efficiency
under certain conditions \cite{blankenship_text_book}. Due to its fundamental importance, the nanoscale structures and the dynamics of the photosynthetic pigment-protein complexes
(PPCs) have attracted long-standing research interest \cite{grondelle_text_book,neill_review_nat_phys_2013}. In experimental reports on the long-lived quantum coherence in PPCs
\cite{fleming_science_2007,fleming_nature_2007,beating_signals_pnas_2010}, many open questions associated with the presence of robust quantum coherence against the surrounding environment
have been discussed extensively, including its impact on the quantum efficiency \cite{plenio_noise_transport_njp_2008,plenio_noise_transport_jcp_2009,elisabet_quan_coh_nat_phys_2014},
the origin of long-lived quantum coherence \cite{pachon_origin_coherence_jpcl_2011,christensson_origin_coherence_jpcb_2012,plenio_long_coh_jcp_2013,engel_corr_fluc_jcp_2012},
the role played by the surrounding environments \cite{avinash_jcp_2012,plenio_nature_physics_2013}, and the non-Markovian memory effect \cite{hongbin_pre_2014,hongbin_scirep_2015}.

Moreover, these studies also inspire attention from a thermodynamical aspect when considering the PPCs as biological quantum heat engines (QHEs)
\cite{dorfman_bio_qhe_pnas_2013,creatore_bio_qhe_prl_2013}. The typical QHEs, such as the working substance of a laser and the semiconductor photocell, generically possess limited
efficiency subject to the detailed balance between absorption and emission of the pumping radiation \cite{einstein_phys_z_1917}. For examples, Scovil and Schulz-DuBois showed that
the efficiency of the maser is described by a Carnot relation \cite{maser_effi_prl_1959}. Shockley and Queisser showed that the efficiency of a photocell is limited to 33\%, due to
radiative recombination, thermalization, and unabsorbed photons \cite{photocell_effi_jap_1961}.

Many studies have been proposed to enhance the performance of QHEs. One promising approach is to utilize the quantum coherence to yield higher output power
\cite{scully_qphotocell_prl_2010,scully_qhe_pnas_2011,svidzinsky_fano_int_pra_2011}. Recently, Dorfman \textit{et al}. \cite{dorfman_bio_qhe_pnas_2013} have proposed that the
noise-induced coherence observed in photosynthetic reaction centers (RCs) is helpful for boosting the photocurrent by at least 27\% compared to an equivalent classical photocell.
This enhancement is attributed to the Fano interference \cite{scully_qphotocell_prl_2010,scully_qhe_pnas_2011,svidzinsky_fano_int_pra_2011}, which originates from the coupling of
two levels to the same continuum. This interference effectively eliminates the radiative dissipation and enables the optical systems to violate the detailed balance that sets an
intrinsic upper bound on the efficiency of light-harvesting devices.

Additionally, Creatore \textit{et al}. \cite{creatore_bio_qhe_prl_2013} alternatively utilize the interference between the delocalized states to improve the photocurrent by at least
35\% compared to one with only localized quantum states. Due to the interference, the engine cycling route is significantly redirected, and each transition rate is shown to
be twice stronger than the uncoupled donor case. A further generalization to the case of three dipoles can be found in Ref. \cite{yiteng_pccp_2015}.

In these proposed models, the environments are assumed to be Markovian for simplicity. However, Markovian environments are not capable of maintaining robust quantum coherence,
which plays a crucial role in enhancing the performance of QHEs. In this work, we introduce the coupling to the vibrational mode in PPCs
\cite{christensson_origin_coherence_jpcb_2012,plenio_long_coh_jcp_2013,engel_corr_fluc_jcp_2012,avinash_jcp_2012,plenio_nature_physics_2013,hongbin_pre_2014,grondelle_vib_phonon_jpcb_1997,
grondelle_vib_phonon_jpcb_2000,jordan_vib_enh_eet_jpcb_2011,vivek_pnas_2013,edward_nat_commun_2014}. From previous studies, it is suggested that certain discrete vibrational modes
coupled to the electronic excitation of the RC cofactors should be treated on the same footing as the RC system itself. Models with non-Markovian coupling can explain the unusually
long-lived quantum coherence observed in PPCs and can reveal enhanced transport properties.

\begin{figure*}
\includegraphics[width=\textwidth]{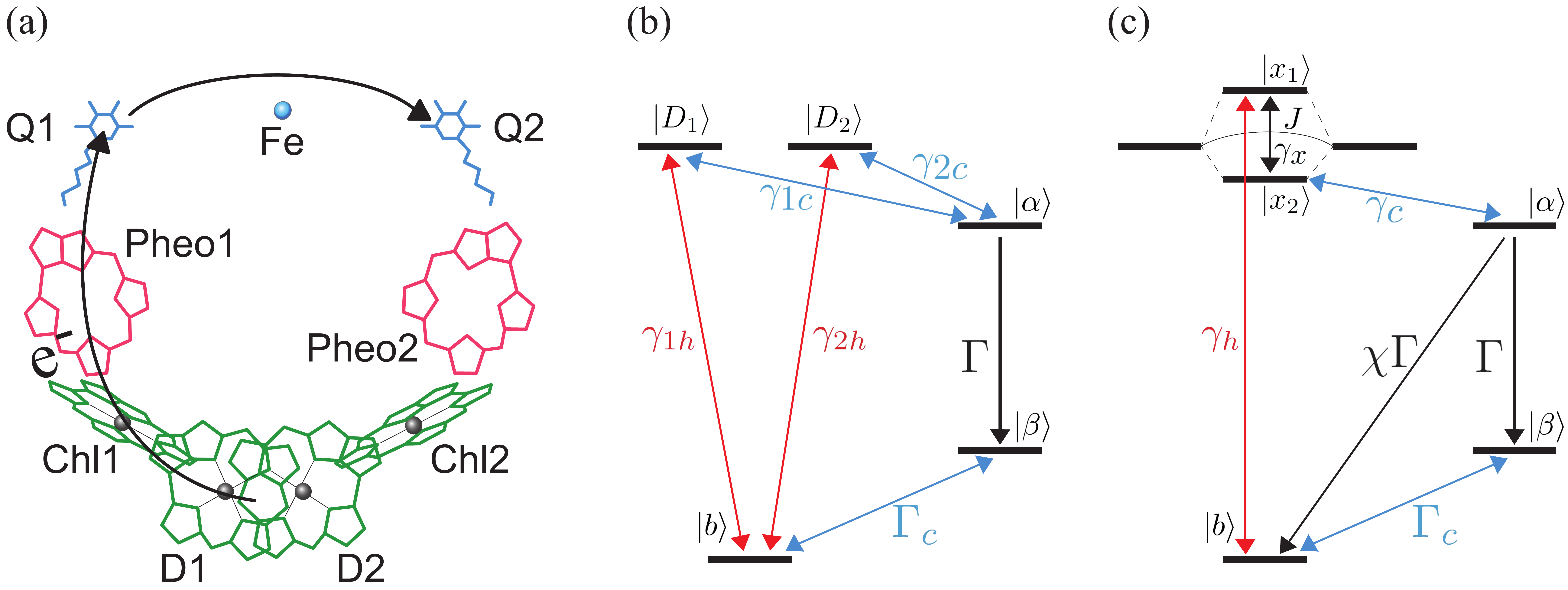}
\caption{(Color online) (a) A reaction center (RC) mainly consists of a chlorophyll dimer (D), two accessory chlorophyll (Chl) molecules, two accessory pheophytin (Pheo) molecules,
and two plastoquinone (Q) molecules. The excited electron is released by the dimer and transferred to the Q2 molecule via the path specified by the black arrows. This process is referred
to as charge separation. (b) The five-level scheme adopted in Ref.~\cite{dorfman_bio_qhe_pnas_2013} is used to simulate a quantum heat engine (QHE) inspired by the RC charge separation
cycle. The quantum coherence in the RC dimer is considered to be able to break the detailed balance and thus enhance the efficiency of the QHE. (c) The refined model including the
dipole-dipole interaction~\cite{creatore_bio_qhe_prl_2013}. The delocalized bright ($|x_1\rangle$) and dark ($|x_2\rangle$) states pave a new efficient route for the engine cycle.
In addition, the energy loss channel via the electron-hole recombination with the rate $\chi\Gamma$ is also taken into account.
}\label{fig_illustration}
\end{figure*}

Our results show that the photocurrent and peak delivered power can be greatly enhanced up to 65 and 63\% with elaborate parameters, respectively. Besides the prominent enhancement
under specific parameters, we also find that the enhancement is more easily achieved for a wide range of parameters. We attribute this improvement to the robust coherence induced
by the non-Markovian vibrational mode, which can modulate the steady-state populations and result in the enhancement of QHE performance.

\section{RC STRUCTURE AND BIOLOGICAL QHE ASPECT}\label{sec_rc_model}

Before explicitly introducing our non-Markovian biological QHE model, it is worthwhile to discuss in detail how a photosynthetic process can be viewed as a cyclic engine model. An
RC typically consists of a pair of chlorophyll molecules, which form an electronically coupled dimer (D), two accessory chlorophyll (Chl) molecules, two accessory pheophytin (Pheo)
molecules, and two plastoquinone (Q) molecules, arranged in two branches associated to the D1 and D2 protein scaffold \cite{bernhard_ps2_org_nature_2005,jan_ps2_org_photosynth_res_2007,mueh_ps2_org_ppb_2008,roberta_ps2_org_jppb_2011}, as shown in Fig.~\ref{fig_illustration}(a). Extensive efforts
have been devoted to identify the site where the electron transfer initiates, and the subsequent transfer pathways. There is evidence showing that two main donors significantly
contribute to the electron transfer process under ambient conditions \cite{grondelle_ct_pathway_biophys_j_2007,grondelle_ct_pathway_biochem_2010,cardona_ps2_review_bba_2012}. In our
model, we suppose the dimer D to be the primary electron donor. This is often the case in bacterial RC, whereas some other species of photosystem II RC use a different pathway which
starts at the accessory Chl of the D1 branch.

Each pigment is usually described by a two-level system with a specific Qy transition energy. When the absorbed solar energy is transferred to the RC dimer, the two dimer molecules
would be excited from the ground state $|b\rangle$ to their excited states, $|D_1\rangle$ and $|D_2\rangle$. This transition is an incoherent exciting process described by the
Hamiltonian $\widehat{V}_\mathrm{h}$, which will be shown below. Sequentially, an excited electron is released by the dimer and transferred to the Q2 molecule via a specific path,
shown in Fig.~\ref{fig_illustration}(a), leaving a hole in the dimer. This process is referred to as \textit{charge separation} and is described by the Hamiltonian
$\widehat{V}_\mathrm{c}$ in the following equation. Then, the Q2 molecule will take the excited electron away from RC and form an effective current $\mathrm{I}=e\Gamma\rho_{\alpha\alpha}$.
Finally, a neutral plastoquinone molecule and a de-excited electron will compensate the positively charged RC with a rate $\Gamma_\mathrm{c}$ and close the charge separation cycle.

In Ref.~\cite{dorfman_bio_qhe_pnas_2013}, Dorfman \textit{et al}. simulate the charge separation cycle by using the five-level scheme shown in Fig.~\ref{fig_illustration}(b). The
free Hamiltonian of this scheme is given by
\begin{eqnarray}
\widehat{H}_\mathrm{RC}&=&\sum_{j = 1,2}\hbar\omega_j|D_j\rangle\langle D_j| \nonumber\\
&&+\hbar\omega_\mathrm{b}|b\rangle\langle b|
+\hbar\omega_\mathrm{\alpha}|\alpha\rangle\langle \alpha|+\hbar\omega_\mathrm{\beta}|\beta\rangle\langle \beta|,
\end{eqnarray}
where $|\alpha\rangle$ is the charge-separated state and $|\beta\rangle$ is the positively charged RC state with a hole in the dimer. Together with the dipole and rotating-wave
approximations, the dimer-reservoir interactions with hot (radiation) and cold (ambient phonon) reservoirs, corresponding to the incoherent excitation and charge separation,
are given by
\begin{equation}
\widehat{V}_\mathrm{h}=\sum_{j=1,2}\sum_\mathbf{k}\hbar\left(g^{(\mathrm{h})}_{j,\mathbf{k}}\hat{\sigma}_{\mathrm{b},j}\otimes\hat{h}^\dagger_\mathbf{k}
+g^{(\mathrm{h})\ast}_{j,\mathbf{k}}\hat{\sigma}^\dagger_{\mathrm{b},j}\otimes\hat{h}_\mathbf{k}\right)
\end{equation}
and
\begin{equation}
\widehat{V}_\mathrm{c}=\sum_{j=1,2}\sum_\mathbf{k}\hbar\left(g^{(\mathrm{c})}_{j,\mathbf{k}}\hat{\sigma}_{\mathrm{\alpha},j}\otimes\hat{c}^\dagger_\mathbf{k}
+g^{(\mathrm{c})\ast}_{j,\mathbf{k}}\hat{\sigma}^\dagger_{\mathrm{\alpha},j}\otimes\hat{c}_\mathbf{k}\right),
\end{equation}
where $g^{(\mathrm{h})}_{j,\mathbf{k}}$ ($g^{(\mathrm{c})}_{j,\mathbf{k}}$) is the coupling strength of $j$th pigment to hot (cold) reservoir mode $\mathbf{k}$,
$\hat{\sigma}_{\mathrm{b},j}=|\mathrm{b}\rangle\langle D_j|$, $\hat{\sigma}_{\mathrm{\alpha},j}=|\mathrm{\alpha}\rangle\langle D_j|$, and $\hat{h}^\dagger_\mathbf{k}$ and
$\hat{c}^\dagger_\mathbf{k}$ ($\hat{h}_\mathbf{k}$ and $\hat{c}_\mathbf{k}$) are the creation (annihilation) operations of the hot and cold reservoirs with wave vector
$\mathbf{k}$, respectively.

Invoking the conventional second-order perturbative treatment with respect to $\widehat{V}_\mathrm{h}$ and $\widehat{V}_\mathrm{c}$, the Born-Markov approximation, and the
Weisskopf-Wigner approximation, the effect of the reservoirs can be described by the Lindblad-type superoperators:
\begin{eqnarray}
&&\mathfrak{R}_{jk\mathrm{h(c)}}\left\{\rho\right\}=\frac{\gamma_{jk\mathrm{h(c)}}}{2}\left[\left(\bar{n}_{j\mathrm{h(c)}}+1\right)
\left(\hat{\sigma}_{\mathrm{b(\alpha)},j}\rho\hat{\sigma}^\dagger_{\mathrm{b(\alpha)},k}\right.\right.    \nonumber\\
&&\left.+\hat{\sigma}_{\mathrm{b(\alpha)},k}\rho\hat{\sigma}^\dagger_{\mathrm{b(\alpha)},j}-\hat{\sigma}^\dagger_{\mathrm{b(\alpha)},k}\hat{\sigma}_{\mathrm{b(\alpha)},j}\rho
-\rho\hat{\sigma}^\dagger_{\mathrm{b(\alpha)},j}\hat{\sigma}_{\mathrm{b(\alpha)},k}\right)                 \nonumber\\
&&+\bar{n}_{j\mathrm{h(c)}}\left(\hat{\sigma}^\dagger_{\mathrm{b(\alpha)},j}\rho\hat{\sigma}_{\mathrm{b(\alpha)},k}
+\hat{\sigma}^\dagger_{\mathrm{b(\alpha)},k}\rho\hat{\sigma}_{\mathrm{b(\alpha)},j}\right.                 \nonumber\\
&&\left.\left.-\hat{\sigma}_{\mathrm{b(\alpha)},k}\hat{\sigma}^\dagger_{\mathrm{b(\alpha)},j}\rho
-\rho\hat{\sigma}_{\mathrm{b(\alpha)},k}\hat{\sigma}^\dagger_{\mathrm{b(\alpha)},j}\right)\right],
\label{eq_mark_dipp}
\end{eqnarray}
where $\bar{n}_{j\mathrm{h}}$ are the photon occupations of the hot radiation reservoir;
$\bar{n}_{j\mathrm{c}}=\left[\exp(\hbar(\omega_j-\omega_\mathrm{\alpha})/k_\mathrm{B}T_\mathrm{a})-1\right]^{-1}$ are the average phonon number at ambient temperature $T_\mathrm{a}$;
$\gamma_{jj\mathrm{h(c)}}=\gamma_{j\mathrm{h(c)}}$ are the decay rates from $|D_j\rangle$ to $|b\rangle(|\mathrm{\alpha}\rangle)$ shown in Fig.~\ref{fig_illustration}(b), respectively;
$\gamma_{12\mathrm{h}(\mathrm{c})}=\gamma_{21\mathrm{h}(\mathrm{c})}$ is the cross-couplings describing the effect of interference with
$\gamma_{12\mathrm{h}(\mathrm{c})}=\sqrt{\gamma_{1\mathrm{h}(\mathrm{c})}\gamma_{2\mathrm{h}(\mathrm{c})}}$ for the fully coherent model and $\gamma_{12\mathrm{h}(\mathrm{c})}=0$ for
the incoherent case.

Creatore \textit{et al}. \cite{creatore_bio_qhe_prl_2013} further refined this model by considering the dipole-dipole interaction between the dimer chlorophyll molecules:
\begin{equation}
\widehat{H}_{\mathrm{J}}=J\left(\hat{p}+\hat{p}^\dagger\right), \label{eq_dip_dip_int}
\end{equation}
where $\hat{p}=|D_1\rangle\langle D_2|$ is the dipole operator. Due to the dipole-dipole interaction, the eigenstates of the dimer become the delocalized states $|x_1\rangle$ and $|x_2\rangle$,
as shown in Fig.~\ref{fig_illustration}(c).

Crucially, $|x_1\rangle$ is a symmetric combination and $|x_2\rangle$ is characterized by a relative phase of $\pi$ in the superposition of the localized states. They are referred
to as bright and dark states, respectively, due to their different optical transition properties. The optical transition rate $\gamma_h$ from $b$ to $x_1$ is twice stronger than in the
uncoupled dimer case, whereas the optical transition to $x_2$ is forbidden. Moreover, the electron transition rate $\gamma_c$ from $x_2$ to $\alpha$ is also twice stronger. Namely,
the two delocalized eigenstates pave a new efficient route for the engine cycle as shown in Fig.~\ref{fig_illustration}(c). In addition, the charge-separated state $|\alpha\rangle$
may lose its energy via electron-hole recombination and decay back to the ground state with a rate $\chi\Gamma$.

\section{NON-MARKOVIAN BIOLOGICAL QHE MODEL}

\subsection{Model}

\begin{figure}[th]
\includegraphics[width=\columnwidth]{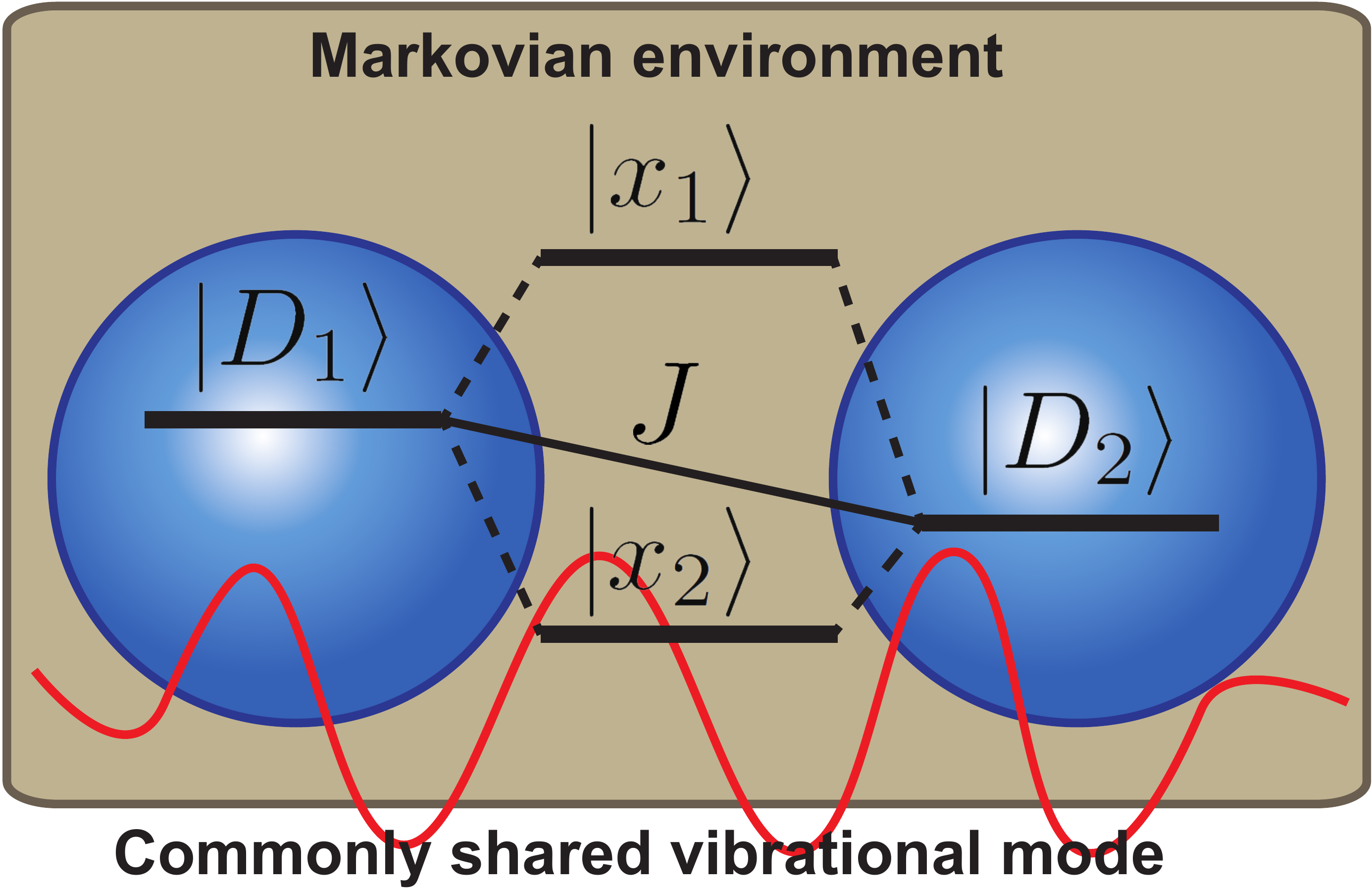}
\caption{(Color online) Schematic illustration of our dimer model. Besides the dipole-dipole interaction, the crucial element of our model is the vibrational modes commonly
shared by the two dimer chlorophyll molecules. This dimer-vibration interaction is treated non-Markovianly and helps resist the decoherence from the Markovian environment.
}\label{fig_dimer_model}
\end{figure}

Based on the molecular structure and the charge separation mechanism introduced in Sec.~\ref{sec_rc_model}, we propose a non-Markovian QHE model which takes into account the
experimentally verified coupling to the discrete vibrational modes \cite{grondelle_vib_phonon_jpcb_1997,grondelle_vib_phonon_jpcb_2000} in PPCs. Motivated by the experimental
reports that the highly correlated environment is responsible for the long-lived quantum coherence and non-Markovian behavior \cite{fleming_science_2007,engel_corr_fluc_jcp_2012},
we specifically consider the vibrational mode $\widehat{H}_\mathrm{vib}=\hbar\omega_{\mathbf{q}}\hat{a}^\dagger_\mathbf{q}\hat{a}_\mathbf{q}$ possessing the frequency
$\omega_\mathbf{q}$ resonant with dimer detuning $\Delta\omega=\omega_1-\omega_2$. The two dimer chlorophyll molecules share this common vibrational mode via the interaction Hamiltonian
\begin{equation}
\widehat{V}_\mathrm{vib}=\sum_{j=1,2}|D_j\rangle\langle D_j|\otimes\hbar\left(g_{j,\mathbf{q}}\hat{a}^\dagger_\mathbf{q}+g^\ast_{j,\mathbf{q}}\hat{a}_\mathbf{q}\right),
\label{eq_dim_vib_int}
\end{equation}
where $g_{j,\mathbf{q}}$ is the coupling strength and $\hat{a}_\mathbf{q}^\dagger$ ($\hat{a}_\mathbf{q}$) is the creation (annihilation) operator for the shared vibrational mode
with wave vector $\mathbf{q}$. Note that we model the vibrational mode based on the fact of narrow sharp peaks in the spectral density function
\cite{grondelle_vib_phonon_jpcb_1997,grondelle_vib_phonon_jpcb_2000}, which leads to long correlation time and implies the underdamped nature of the vibrational mode.

It is critical to notice that the nature of correlated fluctuations suggested by two-dimensional (2D) spectroscopy may be different from the commonly shared vibrational mode. However, an accurate
atomic description of the correlated fluctuations in the environment is quite challenging. Besides, the purpose of this work is not aiming at precisely simulating RC dynamics but
qualitatively the impact of the vibrational mode on the efficiency of QHE. We therefore adopt the simplest way to reproduce the high cross-correlation, by incorporating a common
vibrational structure to mimic the correlated environmental fluctuations. Further discussions on this point and how the common vibrational mode can give high cross-correlation
can be found in Appendix~\ref{app_corr_fluc_and_comm_mode}. Hereafter, the commonly shared vibrational mode will be renamed as \textit{common mode} for brevity.

Although the adequacy and functionality of the vibrational motion are still under debate \cite{fujihashi_nuclear_vib_jcp_2015},
a similar model with common mode is adopted for simulating the oscillations in 2D spectroscopy \cite{christensson_origin_coherence_jpcb_2012}
and the effect of the individual vibrational mode has been taken into account in a recent photosynthetic QHE model \cite{planio_vib_bqhe_jcp_2015}.

Our dimer model is shown schematically in Fig.~\ref{fig_dimer_model}. In addition to the Markovian environment played by the physiological surrounding, the underdamped common
mode is the new element in our model. In order to catch the non-Markovian feature, our model puts the common mode on the same footing as the RC itself and treats the interaction
in a non-perturbative manner. Although the Markovian environment is harmful to the quantum coherence, the long correlation time and the non-Markovian effects of the underdamped
common mode can prolong the coherence time of the dimer and help the QHE model resist the decoherence from the Markovian environment.

\subsection{Polaron transformation}

Generically, the first step in dealing with the interaction is to perform the unitary polaron transformation \cite{seogjoo_jang_polaron-1,nazir_prb_2011,silbey_vari._pola._transf._ptrsa_2012}
with respect to the common mode $\hat{a}_\mathbf{q}$: $\widetilde{H}=e^{\widehat{S}}\widehat{H}e^{-\widehat{S}}$, where $\widehat{S}=\sum_{j=1,2}|D_j\rangle\langle D_j|\otimes\left(\frac{g_{j,\mathbf{q}}}{\omega_\mathbf{q}}\hat{a}^\dagger_\mathbf{q}-\frac{g^\ast_{j,\mathbf{q}}}{\omega_\mathbf{q}}\hat{a}_\mathbf{q}\right)$.
The transformed free Hamiltonian reads
\begin{eqnarray}
\widetilde{H}_\mathrm{RC}&=&\sum_{j = 1,2}\hbar\left(\omega_j-\frac{|g_{j,\mathbf{q}}|^2}{\omega_\mathbf{q}}\right)|D_j\rangle\langle D_j| \nonumber\\
&&+\hbar\omega_\mathrm{b}|b\rangle\langle b|
+\hbar\omega_\mathrm{\alpha}|\alpha\rangle\langle \alpha|+\hbar\omega_\mathrm{\beta}|\beta\rangle\langle \beta|. \label{eq_p_tran_rc_hami}
\end{eqnarray}
The first term on the right-hand side of Eq.~(\ref{eq_p_tran_rc_hami}) includes the reorganization energy given by the common mode. After transformation, $\widehat{H}_\mathrm{vib}$
remains intact and $\widehat{V}_\mathrm{vib}$ is formally eliminated. This does not imply that the dimer-vibration interaction is canceled. Instead, it is depicted by the transformed
dipole-dipole interaction:
\begin{equation}
\widetilde{H}_\mathrm{J}=J\left(\hat{p}\otimes\widehat{X}+\hat{p}^\dagger\otimes\widehat{X}^\dagger\right), \label{eq_p_tran_dip_dip}
\end{equation}
where $\widehat{X}=\widehat{D}_1\widehat{D}^\dagger_2$, and $\widehat{D}_j=\exp\left[\frac{g_{j,\mathbf{q}}}{\omega_\mathbf{q}}\hat{a}^\dagger_\mathbf{q}
-\frac{g^\ast_{j,\mathbf{q}}}{\omega_\mathbf{q}}\hat{a}_\mathbf{q}\right]$ is the displacement operator.

\subsection{Master equation for the RC-vibration joint system}

Akin to the previous works, the dimer-reservoir interactions $\widehat{V}_\mathrm{h}$ and $\widehat{V}_\mathrm{c}$ are treated perturbatively.
This results in the Lindblad-type superoperators shown in Eqs.~(\ref{eq_mark_dipp}). In addition to the RC electronic degrees of freedom,
we consider the RC-vibration joint master equation in the interaction picture with respect to $\widetilde{H}_\mathrm{RC}+\widehat{H}_\mathrm{vib}$
\begin{eqnarray}
\frac{\partial}{\partial t}\tilde{\rho}(t)&=&-\frac{i}{\hbar}\left[\widetilde{H}_\mathrm{J}(t),\tilde{\rho}(t)\right]                 \nonumber\\
&&+\sum_{j=1,2}\mathfrak{R}_{jk\mathrm{h}}\left\{\tilde{\rho}(t)\right\}+\mathfrak{R}_{jk\mathrm{c}}\left\{\tilde{\rho}(t)\right\}   \nonumber\\
&&+\mathfrak{R}_\mathrm{\Gamma}\left\{\tilde{\rho}(t)\right\}+\mathfrak{R}_{\mathrm{\Gamma}_\mathrm{c}}\left\{\tilde{\rho}(t)\right\}
+\mathfrak{R}_\mathrm{\chi\Gamma}\left\{\tilde{\rho}(t)\right\}.          \nonumber\\     \label{eq_dim_lind_form}
\end{eqnarray}
The three additional Lindblad-type superoperators
\begin{equation}
\mathfrak{R}_\mathrm{\Gamma}\left\{\tilde{\rho}(t)\right\}=
\Gamma\left[\hat{\sigma}_{\mathrm{\beta},\mathrm{\alpha}}\tilde{\rho}(t)\hat{\sigma}^\dagger_{\mathrm{\beta},\mathrm{\alpha}}
-\frac{1}{2}\left\{\hat{\sigma}^\dagger_{\mathrm{\beta},\mathrm{\alpha}}\hat{\sigma}_{\mathrm{\beta},\mathrm{\alpha}},\tilde{\rho}(t)\right\}\right],
\end{equation}
\begin{eqnarray}
&&\mathfrak{R}_{\mathrm{\Gamma}_\mathrm{c}}\left\{\tilde{\rho}(t)\right\}=   \nonumber\\
&&\Gamma_\mathrm{c}\left(\bar{N}_\mathrm{c}+1\right)\left[\hat{\sigma}_{\mathrm{b},\mathrm{\beta}}\tilde{\rho}(t)\hat{\sigma}^\dagger_{\mathrm{b},\mathrm{\beta}}
-\frac{1}{2}\left\{\hat{\sigma}^\dagger_{\mathrm{b},\mathrm{\beta}}\hat{\sigma}_{\mathrm{b},\mathrm{\beta}},\tilde{\rho}(t)\right\}\right]                             \nonumber\\
&&+\Gamma_\mathrm{c}\bar{N}_\mathrm{c}\left[\hat{\sigma}_{\mathrm{\beta},\mathrm{b}}\tilde{\rho}(t)\hat{\sigma}^\dagger_{\mathrm{\beta},\mathrm{b}}
-\frac{1}{2}\left\{\hat{\sigma}^\dagger_{\mathrm{\beta},\mathrm{b}}\hat{\sigma}_{\mathrm{\beta},\mathrm{b}},\tilde{\rho}(t)\right\}\right],
\end{eqnarray}
and
\begin{equation}
\mathfrak{R}_\mathrm{\chi\Gamma}\left\{\tilde{\rho}(t)\right\}=
\chi\Gamma\left[\hat{\sigma}_{\mathrm{b},\mathrm{\alpha}}\tilde{\rho}(t)\hat{\sigma}^\dagger_{\mathrm{b},\mathrm{\alpha}}
-\frac{1}{2}\left\{\hat{\sigma}^\dagger_{\mathrm{b},\mathrm{\alpha}}\hat{\sigma}_{\mathrm{b},\mathrm{\alpha}},\tilde{\rho}(t)\right\}\right]
\end{equation}
represent the transition $\alpha\rightarrow\beta$, $\beta\leftrightarrow b$, and recombination $\alpha\rightarrow b$, respectively. The advantage of the Lindblad prescription is that
it enforces the coherence to evolve in a physically consistent way. Here, we stress that even if the joint system is governed by a Lindblad-type master equation~(\ref{eq_dim_lind_form})
and seemingly Markovian, this is not the case if we consider the RC reduced dynamics by tracing out the common mode. Non-Markovianity could be induced due to the dimer-vibration interaction.
For further examples on the non-Markovian dynamics of a subsystem out of a Markovian total system, please see Refs.~\cite{apollaro_comparison_pra_2014,hongbin_k_div_diag_pra_2015}.

It should be noted that, compared with Dorfman's model, introducing the interaction to the common mode would modify the Hamiltonian eigenstates, whereas we assume that the non-unitary
part of the master equation~(\ref{eq_dim_lind_form}) is not significantly altered and inherits the same Lindblad-type superoperators (up to a corresponding polaron transformation) as the
one without dimer-vibration coupling. This prescription is shown to be accurate provided two conditions are satisfied \cite{piilo_jc_justification_pra_2007,piilo_jc_justification_jpa_2007}.
One is that the Hamiltonian transition frequencies are much larger than the decay rate, and the other is that the spectrum of the surrounding Markovian environment is relative flat and
featureless. These two conditions are met in our model. This well justifies the usage of the Lindblad prescription.

\section{STEADY-STATE CURRENT AND POWER ENHANCEMENT}

We develop an analytical method to solve the steady-state solutions of the master equation~(\ref{eq_dim_lind_form}). It should be noted that, in solving the steady-state solutions,
we introduce a decoherence channel phenomenologically with the decoherence time $\tau_2$ to simulate the effect of physiological surroundings. Details can be found in the Appendices.
As stated in Sec.~\ref{sec_rc_model}, the current is formed by taking the excited electron away. The corresponding voltage is defined as the chemical potential difference between
the two states $\mathrm{\alpha}$ and $\mathrm{\beta}$, $eV\equiv\mu_\mathrm{\alpha}-\mu_\mathrm{\beta}$. By using the Boltzmann distribution,
$\rho_{jj}\propto\exp\left[-\left(\hbar\omega_j-\mu_j\right)/k_\mathrm{B}T_\mathrm{a}\right]$, the output
voltage can be expressed in terms of the RC steady-state population:
\begin{equation}
eV=\hbar\omega_\mathrm{\alpha}-\hbar\omega_\mathrm{\beta}+k_\mathrm{B}T_\mathrm{a}\ln\left(\frac{\rho_\mathrm{\alpha\alpha}}{\rho_\mathrm{\beta\beta}}\right).
\end{equation}

\begin{table}
\caption{Parameters used in the calculations.}
\begin{tabular}{c| c c c c}
\hline\hline
                                     & Figure~\ref{fig_current_enhancement} & Figure~\ref{fig_steady_state} & Figure~\ref{fig_cur_enh_var_J} & Figure~\ref{fig_power_enhancement} \\
\hline
$\omega_1$ (cm$^{-1}$)               & 14856 & 14856 & 14856 & 14856 \\
$\omega_2$ (cm$^{-1}$)               & 14756 & 14756 & 14756 & 14756 \\
$\omega_\mathrm{\alpha}$ (cm$^{-1}$) & 13205 & 13205 & 13205 & 13205 \\
$\omega_\mathrm{\beta}$ (cm$^{-1}$)  & 1651  & 1651  & 1651  & 1651  \\
$\omega_\mathrm{b}$ (cm$^{-1}$)      & 0     & 0     & 0     & 0     \\
$\mathrm{J}$ (cm$^{-1}$)             & 100   & 100   &Varying& 100   \\
\hline
$\omega_\mathbf{q}$ (cm$^{-1}$)      & 100   & 100   & 100   & 100   \\
$\mathrm{S}_{1,\mathbf{q}}$ (10$^{-3}$)& 0.3 & 0.3   & 0.3   & 0.3   \\
$\mathrm{S}_{2,\mathbf{q}}$ (10$^{-3}$)&Varying&Varying&Varying&2\\
\hline
$\gamma_{1\mathrm{h}}$ (cm$^{-1}$)   & 0.005 & 0.005 & 0.005 & 0.005 \\
$\gamma_{2\mathrm{h}}$ (cm$^{-1}$)   & 0.0016& 0.0016& 0.0016& 0.0016\\
$\gamma_\mathrm{c}$    (cm$^{-1}$)   & 158   & 158   & 158   & 158   \\
\hline
$\bar{n}_{1\mathrm{h}}$              & 60000 & 60000 & 60000 & 60000 \\
$\bar{n}_{2\mathrm{h}}$              & 10000 & 10000 & 10000 & 10000 \\
$T_\mathrm{a}$ ($\mathrm{K}$)        & 300   & 300   & 300   & 300   \\
$1/\tau_2$ (cm$^{-1}$)               & 41    & 41    & 41    & 41    \\
\hline
$\Gamma$ (cm$^{-1}$)                 & 1000  & 1000  & 1000  &Varying\\
$\Gamma_\mathrm{c}$ (cm$^{-1}$)      & 200   & 200   & 200   & 200 \\
$\chi$                               & 20\%  & 20\%  & 20\%  & 20\% \\
\hline\hline
\end{tabular}
\label{tab_parameters}
\end{table}

All the parameters used in the following calculations are listed in Table~\ref{tab_parameters}. Most of the RC parameters are taken from the overdamped regime in
Ref.~\cite{dorfman_bio_qhe_pnas_2013} with some Qy altered energies, for which the greatest enhancement in the current is up to 27\%, while the coupling strength $g_{j,\mathbf{q}}$
to the common mode in Eq.~(\ref{eq_dim_vib_int}) is usually expressed by an experimentally measurable quantity, the Huang-Rhys factor
($\mathrm{S}_{j,\mathbf{q}}=|g_{j,\mathbf{q}}|^2/\omega^2_\mathbf{q}$). Due to the magnitude of the Huang-Rhys factor used in this work, the reorganization energy in
Eq.~(\ref{eq_p_tran_rc_hami}) is negligibly small compared with the Qy transition energy of each molecule. To investigate the dependence on the individual couplings to the cold reservoir,
we fix $\gamma_\mathrm{c}=\gamma_{1\mathrm{c}}+\gamma_{2\mathrm{c}}$.

\begin{figure}[th]
\includegraphics[width=\columnwidth]{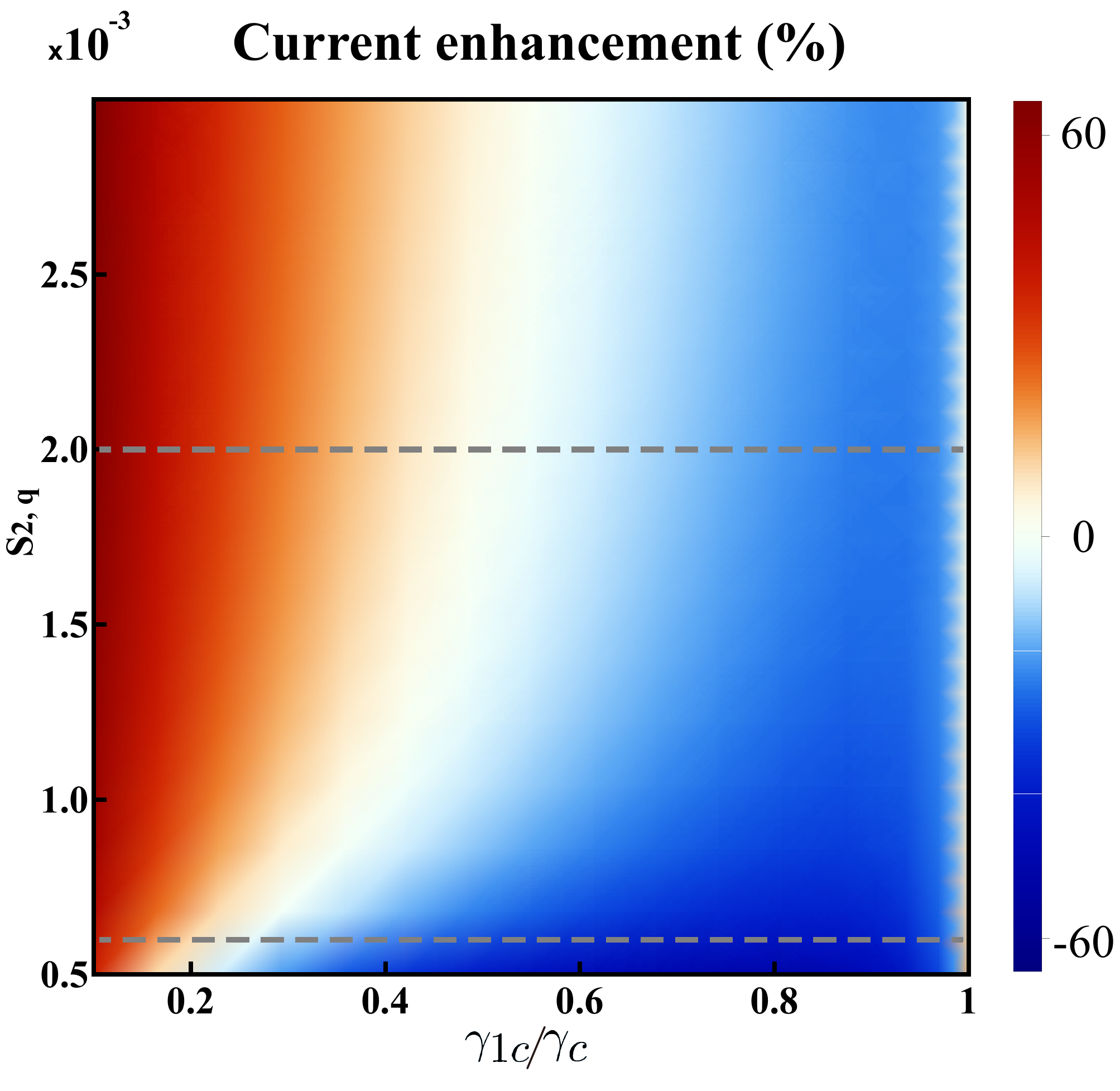}
\caption{(Color online) The current enhancement $(\mathrm{I}-\mathrm{I}_\mathrm{MAR})/\mathrm{I}_\mathrm{MAR}$ as functions of the ratio $\gamma_{1\mathrm{c}}/\gamma_\mathrm{c}$
and the Huang-Rhys factor $\mathrm{S}_{2,\mathbf{q}}$. The coupling to the common mode mainly gives the enhancement if $\gamma_{1\mathrm{c}}$ is weak, whereas the current is
suppressed if $\gamma_{1\mathrm{c}}$ is strong. The two horizontal gray dashed lines denote $\mathrm{S}_{2,\mathbf{q}}=0.002$ and $0.0006$, respectively.
}\label{fig_current_enhancement}
\end{figure}

In Fig.~\ref{fig_current_enhancement}, we show the relative current enhancement $(\mathrm{I}-\mathrm{I}_\mathrm{MAR})/\mathrm{I}_\mathrm{MAR}$, where $\mathrm{I}$ ($\mathrm{I}_\mathrm{MAR}$)
is the steady-state current of our QHE model (the incoherent Markovian model). When $\mathrm{S}_{2,\mathbf{q}}$ is small, the steady-state current is mostly suppressed in the presence
of the common mode. However, when $\mathrm{S}_{2,\mathbf{q}}$ increases, the common mode can significantly enhance the current by up to 65\% if $\gamma_{1\mathrm{c}}$ is weak, whereas
the current is still suppressed if $\gamma_{1\mathrm{c}}$ is strong. Similar results can also be found in Refs.~\cite{dorfman_bio_qhe_pnas_2013,creatore_bio_qhe_prl_2013}, where the
performance of QHEs is not always enhanced even in the presence of coherence. To understand the underlying reason of these results, it deserves deeper investigations. Since the current
is proportional to the population, it is intriguing to see how the steady-state population changes with the parameters.

\begin{figure}[th]
\includegraphics[width=\columnwidth]{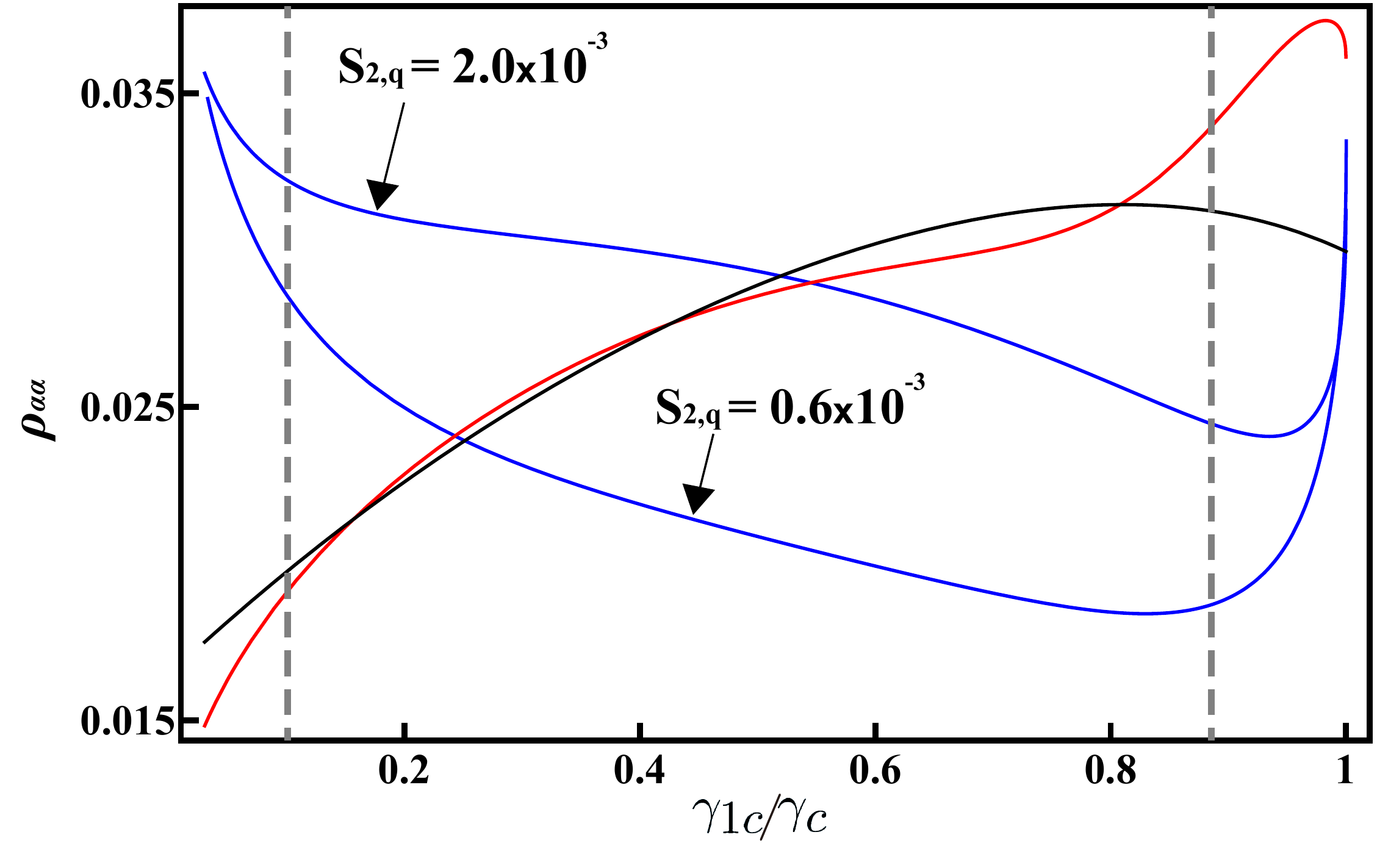}
\caption{(Color online) The steady-state populations of $\rho_\mathrm{\alpha\alpha}$ as a function of the ratio $\gamma_{1\mathrm{c}}/\gamma_\mathrm{c}$ for fully coherent Markovian
model (red curve), incoherent Markovian model (black curve), and our non-Markovian QHE (blue curves) at $\mathrm{S}_{2,\mathbf{q}}=0.002$ and $0.0006$. Steady-state population profiles
are modulated with increasing $\mathrm{S}_{2,\mathbf{q}}$. The QHE performance can easily be enhanced within a wider range in the presence of the common mode, even if $\gamma_{1\mathrm{c}}$
is slightly less than $\gamma_{2\mathrm{c}}$, while quantum coherence can increase the steady-state population in the Markovian model only when $\gamma_{1\mathrm{c}}$ is much larger than
$\gamma_{2\mathrm{c}}$.
}\label{fig_steady_state}
\end{figure}

To illustrate this, in Fig.~\ref{fig_steady_state}, we compare $\rho_\mathrm{\alpha\alpha}$ when the RC reaches the steady-state operation of the fully coherent Markovian model
(red curve), the incoherent Markovian model (black curve), and our QHE (blue curves) for $\mathrm{S}_{2,\mathbf{q}}=0.002$ and $0.0006$ (corresponding to the two gray dashed lines
in Fig.~\ref{fig_current_enhancement}). It can be seen that, comparing both Markovian models, the coherence can result in significant enhancement only when $\gamma_{1\mathrm{c}}$
is much larger than $\gamma_{2\mathrm{c}}$. The strong enhancement obtained in Ref.~\cite{dorfman_bio_qhe_pnas_2013} corresponds to the right vertical gray dashed line at
$\gamma_{1\mathrm{c}}/\gamma_\mathrm{c}=0.886$. On the other hand, the population profiles of our model are lifted when increasing $\mathrm{S}_{2,\mathbf{q}}$. One therefore can
obtain an enhancement within a wider range in the presence of the common mode, even if $\gamma_{1\mathrm{c}}$ is slightly less than $\gamma_{2\mathrm{c}}$. This shows that, to achieve the
enhancement, the robustness of quantum coherence is also a key ingredient. Consequently, the common mode is helpful in achieving enhancement. This improvement results from
the long-lived coherence induced by the common mode, which gives rise to the Fano interference \cite{scully_qphotocell_prl_2010,scully_qhe_pnas_2011,svidzinsky_fano_int_pra_2011}
and enhances/suppresses some absorption/emission processes. Therefore, the steady-state populations are modulated.

\begin{figure}[th]
\includegraphics[width=\columnwidth]{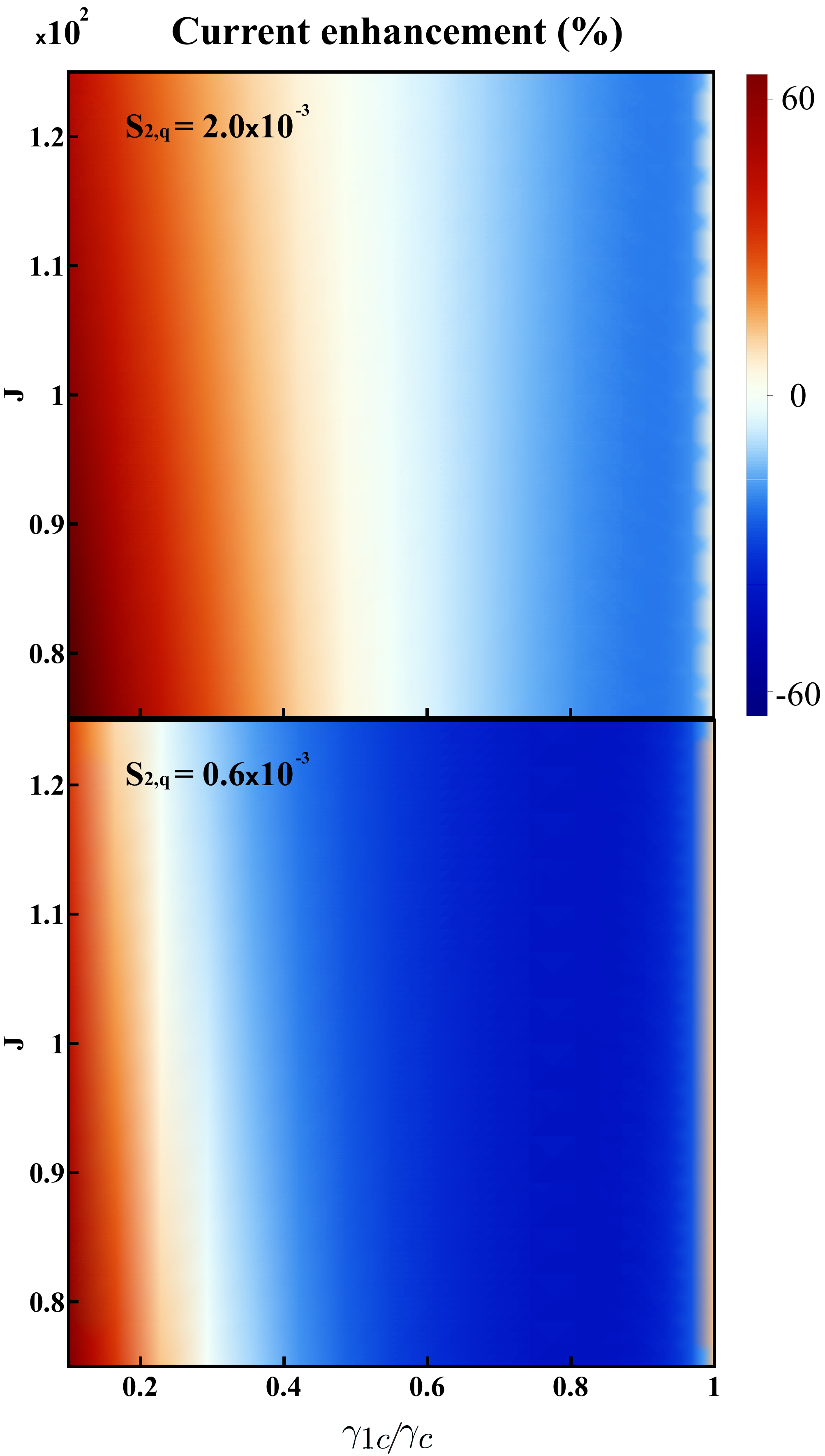}
\caption{(Color online) The current enhancement $(\mathrm{I}-\mathrm{I}_\mathrm{MAR})/\mathrm{I}_\mathrm{MAR}$ as functions of the ratio $\gamma_{1\mathrm{c}}/\gamma_\mathrm{c}$
and the dipole-dipole interaction strength $J$ for $\mathrm{S}_{2,\mathbf{q}}=0.002$ (upper panel) and $0.0006$ (lower panel), corresponding to the two gray dashed lines in
Fig.~\ref{fig_current_enhancement}. The current enhancement does not change significantly with $J$.
}\label{fig_cur_enh_var_J}
\end{figure}

Besides the common mode, the presence of dipole-dipole interaction Eq.~(\ref{eq_dip_dip_int}), resulting in the delocalized states, also alters the properties of the inter-molecular
coherence. To compare the coherence given by two different mechanisms, we also investigate the dependence of relative current enhancement on the dipole-dipole interaction. The results
are shown in Fig.~\ref{fig_cur_enh_var_J} for $\mathrm{S}_{2,\mathbf{q}}=0.002$ and $0.0006$ (corresponding to the two gray dashed lines in Fig.~\ref{fig_current_enhancement}). It can
be seen that the current enhancement is not sensitive to the coupling strength $J$. This reveals the different utilities of the two mechanisms. The effect of dipole-dipole interaction
is mainly governed in the master equation (\ref{eq_dim_lind_form}) and is not capable of recovering the lost coherence. Therefore, the coherence associated to the dipole-dipole
interaction is fragile under the decoherence process of the Markovian environment. On the other hand, the common mode can prolong the coherence time and has prominent impact on the QHE
performance.

\begin{figure}[th]
\includegraphics[width=\columnwidth]{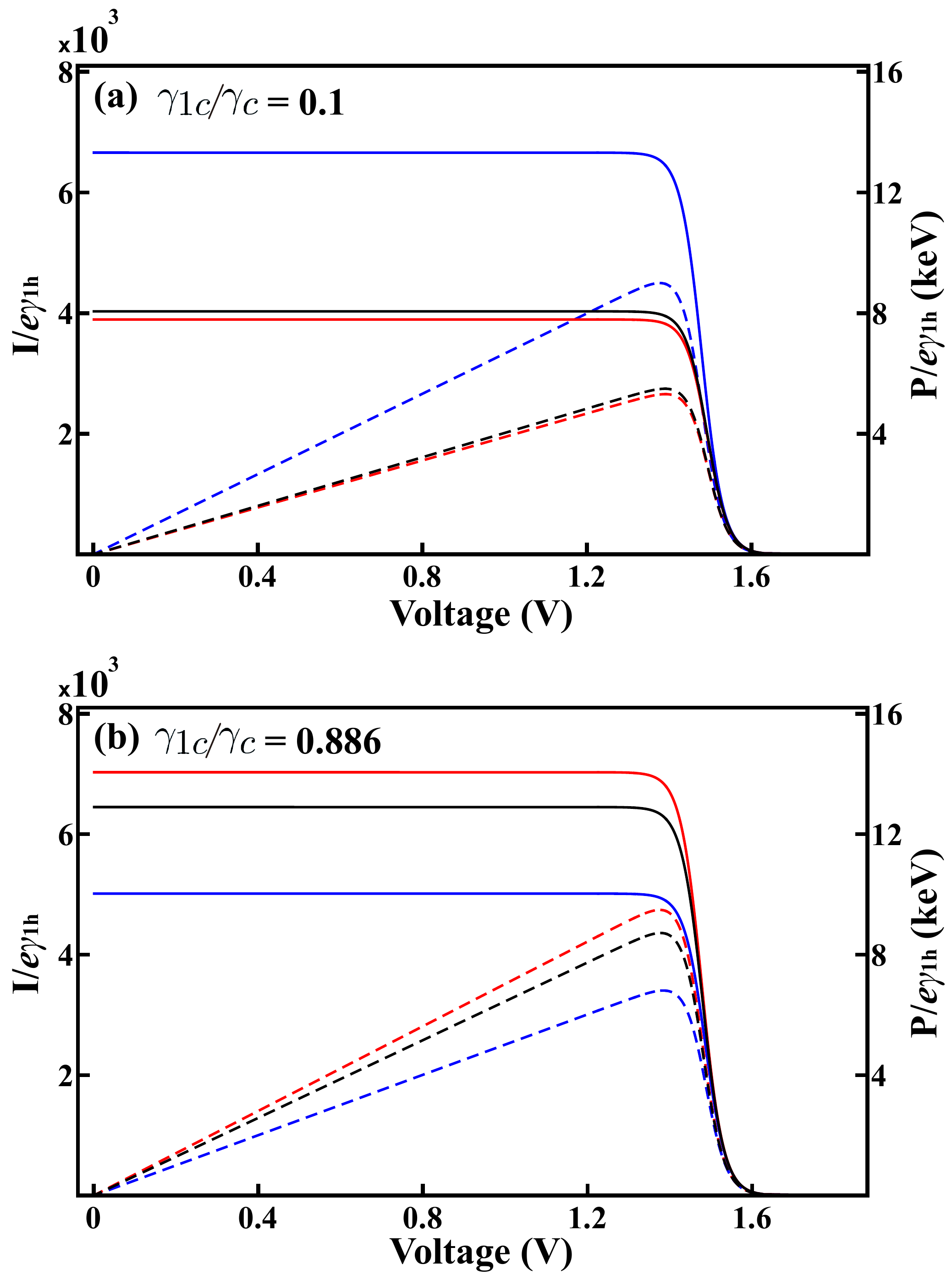}
\caption{(Color online) Steady-state I-V characteristic (solid curves) and power (dashed curves) as a function of voltage for fully coherent Markovian model (red curves), incoherent
Markovian model (black curves), and our non-Markovian QHE (blue curves) at (a) $\gamma_{1\mathrm{c}}/\gamma_\mathrm{c}=0.1$ and (b) $\gamma_{1\mathrm{c}}/\gamma_\mathrm{c}=0.886$.
In (a), the result corresponds to a weaker value of $\gamma_{1\mathrm{c}}$. The current and peak delivered power are enhanced by roughly 65\% and 63\% in the non-Markovian QHE model,
respectively, whereas the performance is slightly suppressed in the coherent Markovian model. In (b), the result corresponds to a stronger value of $\gamma_{1\mathrm{c}}$ and our QHE
model fails to improve the performance.
}\label{fig_power_enhancement}
\end{figure}

The current-voltage (I-V) characteristic in the steady state limit is obtained by varying the rate $\Gamma$, from $\Gamma=0$ to a large $\Gamma$. This corresponds to the transition
from the open circuit regime ($\mathrm{I}=0$ and $\mathrm{V}=\mathrm{V}_\mathrm{OC}$) to the shortcut regime ($\mathrm{I}\rightarrow\mathrm{I}_\mathrm{SC}$ and $\mathrm{V}\rightarrow0$).
The power P is determined by the formula $\mathrm{P}=\mathrm{I}\cdot\mathrm{V}$. Figure~\ref{fig_power_enhancement} shows the current (solid curves) and power (dashed curves) for
three models at $\mathrm{S}_{2,\mathbf{q}}=0.002$, (a) $\gamma_{1\mathrm{c}}/\gamma_\mathrm{c}=0.1$, and (b) $\gamma_{1\mathrm{c}}/\gamma_\mathrm{c}=0.886$, corresponding to the two
vertical gray dashed lines in Fig.~\ref{fig_steady_state}. Figure~\ref{fig_power_enhancement}(a) shows the results of $\gamma_{2\mathrm{c}}>\gamma_{1\mathrm{c}}$. The current is enhanced
by roughly 65\% in our non-Markovian QHE model compared with the incoherent Markovian one, whereas the current and power are slightly suppressed in the coherent Markovian model. And the
resultant peak power is enhanced by about 63\%. On the other hand, Fig.~\ref{fig_power_enhancement}(b) shows the results of $\gamma_{1\mathrm{c}}>\gamma_{2\mathrm{c}}$. The overall
performance of the coherent Markovian model is enhanced due to the presence of coherence; while our non-Markovian QHE model fails to improve the performance.

\section{CONCLUSIONS}

We make a brief survey on the structure and mechanism of the photosynthetic RC. This inspires us to construct a non-Markovian QHE model taking into account the effect of vibrational
modes verified experimentally. To reproduce the high cross-correlation suggested by the 2D spectroscopy experiments, we assume that the vibrational modes are commonly shared
by the dimer in RC. We specifically consider one vibrational mode resonant with the RC dimer and treat it on equal footing as the RC itself.

We first find that in the presence of the common mode the steady-state current can be greatly enhanced by up to 65\% if $\gamma_{2\mathrm{c}}$ is much stronger than $\gamma_{1\mathrm{c}}$.
On the other hand, the current is suppressed if $\gamma_{1\mathrm{c}}$ is stronger or $\mathrm{S}_{2,\mathbf{q}}$ is not strong enough.
To further understand the functionality of the common mode, we investigate how the steady-state population depends on the other parameters. We find that the steady-state population
of our model is lifted with increasing coupling to the common mode. One can easily obtain enhanced QHE performance within a wider range even if $\gamma_{1\mathrm{c}}$ is slightly
less than $\gamma_{2\mathrm{c}}$ in the presence of common mode. We attribute this benefit to the robust coherence induced by the common mode,
which modulates the steady-state population profiles and results in the enhancement of QHE performance.
Furthermore, even if the dipole-dipole interaction is also related to the coherence,
the current enhancement is not sensitive to the coupling strength $J$. Consequently, not only the coherence itself, but also
the robustness is critical for enhancing the QHE performance. These elucidate the utilities of the common mode in QHE.

We finally calculate the steady-state current and peak power delivery for the three models for two different cases. As expected, we conclude that with a stronger value of $\gamma_{2\mathrm{c}}$
the current and the peak delivered power can exceed the incoherent Markovian ones by about 65\% and 63\%, whereas the common mode fails to improve the performance of QHE for the case
of $\gamma_{1\mathrm{c}}>\gamma_{2\mathrm{c}}$.

\section*{ACKNOWLEDGMENTS}

This work is supported partially by the National Center for Theoretical Sciences and Minister of Science and Technology, Taiwan, Grants No. MOST 103-2112-M-006-017-MY4
and No. MOST 104-2811-M-006-059.

\begin{widetext}
\appendix

\section{CORRELATED ENVIRONMENTAL FLUCTUATIONS AND COMMON MODE}\label{app_corr_fluc_and_comm_mode}

As mentioned in the main text, models with correlated environmental fluctuations are often used to simulate the exciton dynamics in 2D-spectroscopy.
The following Hamiltonian is frequently adopted:
\begin{equation}
\widehat{H}_\mathrm{cor}=\sum_{j,k=1,2}|D_j\rangle\langle D_k|\left(\delta_{j,k}(\hbar\omega_j+\widehat{A}_j)+(1-\delta_{j,k})J\right)+\widehat{H}_\mathrm{bath},
\end{equation}
where $\widehat{A}_j$ is the operator on the bath associated to site $|D_j\rangle$. In such a model, the interaction of different modes within the bath is not explicitly
considered. The correlated fluctuations is described via the site correlation and cross-correlation functions
\begin{equation}
C_{j,k}(t)=\langle\widehat{A}_j(t)\widehat{A}_k(0)\rangle,
\end{equation}
where $\widehat{A}_j(t)=\exp\left(i\widehat{H}_\mathrm{bath}t/\hbar\right)\widehat{A}_j\exp\left(-i\widehat{H}_\mathrm{bath}t/\hbar\right)$.
The extent to which the fluctuations associated to each site are correlated is characterized by a parameter $c$:
\begin{equation}
C_{12}=c\sqrt{C_{11}C_{22}}.
\end{equation}
The fluctuations is called to be highly correlated when $c$ approaches $1$.

If we further assume that $\widehat{A}_j=|g_j|\widehat{A}$ and $\widehat{A}=\hat{a}_\mathbf{q}^\dagger+\hat{a}_\mathbf{q}$, it is easy to see that $c=1$ and this corresponds to our
model with real coupling strength $g_{j,\mathbf{q}}$ in Eq.~(\ref{eq_dim_vib_int}). Namely, our model can be considered as the simplest way to reproduce the unital cross-correlation.

\section{SOLUTIONS TO THE RC-VIBRATION JOINT MASTER EQUATION}

Now we present an analytical method to solve the steady-state solutions of Eq.~(\ref{eq_dim_lind_form}) in the Appendices. Performing a formal time integral on both sides, we can obtain
the formal solution to the RC-vibration joint density matrix:
\begin{equation}
\tilde{\rho}(t)=\rho(0)-\frac{i}{\hbar}\int_0^t\left[\widetilde{H}_\mathrm{J}(\tau),\tilde{\rho}(\tau)\right]d\tau
+\int_0^t\sum_\mu\mathfrak{R}_\mu\left\{\tilde{\rho}(\tau)\right\}d\tau,     \label{eq_dim_formal_sol}
\end{equation}
where the last term on the right-hand side is the summation over all Lindblad-type superoperators in Eq.~(\ref{eq_dim_lind_form}). Recall that, in the interaction picture, the expectation
value of a polaron-transformed operator $\widetilde{O}(t)$
\begin{equation}
\langle\widehat{O}\rangle_t=\mathrm{Tr}\left(\rho(t)\widehat{O}\right)=\mathrm{Tr}\left(\tilde{\rho}(t)\widetilde{O}(t)\right)
\end{equation}
is defined as tracing over the RC-vibration joint system. Similarly, the one for the dipole operator reads
\begin{equation}
\langle\hat{p}\rangle_t=\mathrm{Tr}\left(\tilde{\rho}(t)\tilde{p}(t)\right),
\end{equation}
where $\tilde{p}(t)=\hat{p}e^{i\left(\omega_1-\omega_2\right)t}\otimes\widehat{X}(t)$ is the polaron-transformed dipole operator in the interaction picture.

With these definitions, it is convenient to derive the equations of motion for the reduced RC density matrix directly from the formal solution Eq.~(\ref{eq_dim_formal_sol}). Applying
the Born approximation, namely, $\tilde{\rho}(t)=\tilde{\rho}_\mathrm{RC}(t)\otimes\rho_\mathrm{vib}$ with
$\rho_\mathrm{vib}=\exp\left[-\frac{\widehat{H}_\mathrm{vib}}{k_\mathrm{B}T_\mathrm{a}}\right]/Z$ being the thermal state of the common mode, one obtains a set of coupled integral equations straightforwardly via explicitly expanding Eq.~(\ref{eq_dim_formal_sol}) or multiplying Eq.~(\ref{eq_dim_formal_sol}) with $\tilde{p}(t)$ and $\tilde{p}^\dagger(t)$ and performing the
trace:
\begin{eqnarray}
\rho_{11}(t)-\rho_{11}(0)&=&-\frac{i}{\hbar}J\int_0^t\left[\langle\hat{p}\rangle_\tau-\langle\hat{p}^\dagger\rangle_\tau\right]d\tau
-\gamma_{1\mathrm{h}}\int_0^t\left[(\bar{n}_\mathrm{1h}+1)\rho_{11}(\tau)-\bar{n}_\mathrm{1h}\rho_{\mathrm{bb}}(\tau)\right]d\tau                  \nonumber\\
&&-\gamma_{1\mathrm{c}}\int_0^t\left[(\bar{n}_{1\mathrm{c}}+1)\rho_{11}(\tau)-\bar{n}_\mathrm{1c}\rho_\mathrm{\alpha\alpha}(\tau)\right]d\tau
-\left[\frac{\gamma_\mathrm{12h}}{2}(\bar{n}_\mathrm{2h}+1)+\frac{\gamma_\mathrm{12c}}{2}(\bar{n}_\mathrm{2c}+1)\right]\int_0^t\left[\langle\hat{p}\rangle_\tau+\langle\hat{p}^\dagger\rangle_\tau\right]d\tau,\nonumber\\
\rho_{22}(t)-\rho_{22}(0)&=&\frac{i}{\hbar}J\int_0^t\left[\langle\hat{p}\rangle_\tau-\langle\hat{p}^\dagger\rangle_\tau\right]d\tau
-\gamma_\mathrm{2h}\int_0^t\left[(\bar{n}_\mathrm{2h}+1)\rho_{22}(\tau)-\bar{n}_\mathrm{2h}\rho_\mathrm{bb}(\tau)\right]d\tau                      \nonumber\\
&&-\gamma_\mathrm{2c}\int_0^t\left[(\bar{n}_\mathrm{2c}+1)\rho_{22}(\tau)-\bar{n}_\mathrm{2c}\rho_\mathrm{\alpha\alpha}(\tau)\right]d\tau
-\left[\frac{\gamma_\mathrm{12h}}{2}(\bar{n}_\mathrm{1h}+1)+\frac{\gamma_\mathrm{12c}}{2}(\bar{n}_\mathrm{1c}+1)\right]\int_0^t\left[\langle\hat{p}\rangle_\tau+\langle\hat{p}^\dagger\rangle_\tau\right]d\tau,\nonumber\\
\langle\hat{p}\rangle_t-\langle\hat{p}\rangle_0&=&-\frac{i}{\hbar}J\int_0^te^{i(\omega_1-\omega_2)(t-\tau)}\left[\rho_{11}(\tau)C(t-\tau)-\rho_{22}(\tau)C^\ast(t-\tau)\right]d\tau                \nonumber\\
&&-\left[\sum_{j,\mu}\frac{\gamma_{j\mu}}{2}\left(\bar{n}_{j\mu}+1\right)+\frac{1}{\tau_2}\right]\int_0^te^{i(\omega_1-\omega_2)(t-\tau)}\langle\hat{p}\rangle_\tau d\tau                          \nonumber\\
&&-\frac{\gamma_\mathrm{12h}}{2}\int_0^te^{i(\omega_1-\omega_2)(t-\tau)}\left[(\bar{n}_\mathrm{1h}+1)\rho_{11}(\tau)C(t-\tau)+(\bar{n}_\mathrm{2h}+1)\rho_{22}(\tau)C^\ast(t-\tau)\right]d\tau     \nonumber\\
&&+\frac{\gamma_\mathrm{12h}}{2}(\bar{n}_\mathrm{1h}+\bar{n}_\mathrm{2h})\int_0^te^{i(\omega_1-\omega_2)(t-\tau)}\rho_\mathrm{bb}(\tau)C^\prime(t-\tau)d\tau                                       \nonumber\\
&&-\frac{\gamma_\mathrm{12c}}{2}\int_0^te^{i(\omega_1-\omega_2)(t-\tau)}\left[(\bar{n}_\mathrm{1c}+1)\rho_{11}(\tau)C(t-\tau)+(\bar{n}_\mathrm{2c}+1)\rho_{22}(\tau)C^\ast(t-\tau)\right]d\tau     \nonumber\\
&&+\frac{\gamma_\mathrm{12c}}{2}(\bar{n}_\mathrm{1c}+\bar{n}_\mathrm{2c})\int_0^te^{i(\omega_1-\omega_2)(t-\tau)}\rho_\mathrm{\alpha\alpha}(\tau)C^\prime(t-\tau)d\tau,                            \nonumber\\
\langle\hat{p}^\dagger\rangle_t-\langle\hat{p}^\dagger\rangle_0&=&\frac{i}{\hbar}J\int_0^te^{-i(\omega_1-\omega_2)(t-\tau)}\left[\rho_{11}(\tau)C^\ast(t-\tau)-\rho_{22}(\tau)C(t-\tau)\right]d\tau\nonumber\\
&&-\left[\sum_{j,\mu}\frac{\gamma_{j\mu}}{2}\left(\bar{n}_{j\mu}+1\right)+\frac{1}{\tau_2}\right]\int_0^te^{-i(\omega_1-\omega_2)(t-\tau)}\langle\hat{p}^\dagger\rangle_\tau                       \nonumber\\
&&-\frac{\gamma_\mathrm{12h}}{2}\int_0^te^{-i(\omega_1-\omega_2)(t-\tau)}\left[(\bar{n}_\mathrm{1h}+1)\rho_{11}(\tau)C^\ast(t-\tau)+(\bar{n}_\mathrm{2h}+1)\rho_{22}(\tau)C(t-\tau)\right]d\tau    \nonumber\\
&&+\frac{\gamma_\mathrm{12h}}{2}(\bar{n}_\mathrm{1h}+\bar{n}_\mathrm{2h})\int_0^te^{-i(\omega_1-\omega_2)(t-\tau)}\rho_\mathrm{bb}(\tau)C^{\prime\ast}(t-\tau)d\tau                                \nonumber\\
&&-\frac{\gamma_\mathrm{12c}}{2}\int_0^te^{-i(\omega_1-\omega_2)(t-\tau)}\left[(\bar{n}_\mathrm{1c}+1)\rho_{11}(\tau)C^\ast(t-\tau)+(\bar{n}_\mathrm{2c}+1)\rho_{22}(\tau)C(t-\tau)\right]d\tau    \nonumber\\
&&+\frac{\gamma_\mathrm{12c}}{2}(\bar{n}_\mathrm{1c}+\bar{n}_\mathrm{2c})\int_0^te^{-i(\omega_1-\omega_2)(t-\tau)}\rho_\mathrm{\alpha\alpha}(\tau)C^{\prime\ast}(t-\tau)d\tau,                     \nonumber\\
\rho_{\mathrm{\alpha\alpha}}(t)-\rho_{\mathrm{\alpha\alpha}}(0)&=&\sum_{j=1,2}\gamma_{j\mathrm{c}}\int_0^t\left[(\bar{n}_{j\mathrm{c}}+1)\rho_{jj}(\tau)
-\bar{n}_{j\mathrm{c}}\rho_{\mathrm{\alpha\alpha}}(\tau)\right]d\tau-\Gamma(1+\chi)\int_0^t\rho_{\mathrm{\alpha\alpha}}(\tau)d\tau                          \nonumber\\
&&+\frac{\gamma_\mathrm{12c}}{2}(\bar{n}_\mathrm{1c}+1+\bar{n}_\mathrm{2c}+1)\int_0^t\left[\langle\hat{p}\rangle_\tau+\langle\hat{p}^\dagger\rangle_\tau\right]d\tau,                              \nonumber\\
\rho_{\mathrm{\beta\beta}}(t)-\rho_{\mathrm{\beta\beta}}(0)&=&-\Gamma_\mathrm{c}\int_0^t\left[(\bar{N}_\mathrm{c}+1)\rho_{\mathrm{\beta\beta}}(\tau)-\bar{N}_\mathrm{c}\rho_{\mathrm{bb}}(\tau)\right]d\tau
+\Gamma\int_0^t\rho_{\mathrm{\alpha\alpha}}(\tau)d\tau,                                          \nonumber\\
\rho_{\mathrm{bb}}(t)-\rho_{\mathrm{bb}}(0)&=&\sum_{j=1,2}\gamma_{j\mathrm{h}}\int_0^t\left[(\bar{n}_{j\mathrm{h}}+1)\rho_{jj}(\tau)-\bar{n}_{j\mathrm{h}}\rho_{\mathrm{\alpha\alpha}}(\tau)\right]d\tau
+\chi\Gamma\int_0^t\rho_{\mathrm{\alpha\alpha}}(\tau)d\tau \nonumber\\
&&+\Gamma_\mathrm{c}\int_0^t\left[(\bar{N}_\mathrm{c}+1)\rho_{\mathrm{\beta\beta}}(\tau)-\bar{N}_\mathrm{c}\rho_{\mathrm{bb}}(\tau)\right]d\tau
+\frac{\gamma_\mathrm{12h}}{2}(\bar{n}_\mathrm{1h}+1+\bar{n}_\mathrm{2h}+1)\int_0^t\left[\langle\hat{p}\rangle_\tau+\langle\hat{p}^\dagger\rangle_\tau\right]d\tau,          \label{eq_dim_integ_eq}
\end{eqnarray}
where $C(t-\tau)=\mathrm{Tr}\left[\rho_\mathrm{vib}\widehat{X}(t)\widehat{X}^\dagger(\tau)\right]$ and
$C^\prime(t-\tau)=\mathrm{Tr}\left[\rho_\mathrm{vib}\widehat{D}_1^\dagger(\tau)\widehat{X}(t)\widehat{D}_2(\tau)\right]$
are the correlation functions.
For the single common mode in our QHE model, the correlation functions can be calculated as $C(t-\tau)=\exp\left[-\Phi(t-\tau)\right]$ and $C^\prime(t-\tau)=\exp\left[-\Phi^\prime(t-\tau)\right]$ with
\begin{eqnarray}
\Phi(t)&=&\frac{|\Delta g_\mathbf{q}|^2}{\omega^2_\mathbf{q}}\left[\left(1-\cos(\omega_\mathbf{q}t)\right)\coth\left(\frac{\hbar\omega_\mathbf{q}}{2k_\mathrm{B}\mathrm{T}_\mathrm{a}}\right)
+i\sin(\omega_\mathbf{q}t)\right],  \nonumber\\
\Phi^\prime(t)&=&\frac{|\Delta g_\mathbf{q}|^2}{\omega^2_\mathbf{q}}\left(1-\cos(\omega_\mathbf{q}t)\right)\coth\left(\frac{\hbar\omega_\mathbf{q}}{2k_\mathrm{B}\mathrm{T}_\mathrm{a}}\right)
-i\frac{|g_{1,\mathbf{q}}|^2-|g_{2,\mathbf{q}}|^2}{\omega^2_\mathbf{q}}\sin(\omega_\mathbf{q}t),
\end{eqnarray}
with $\Delta g_\mathbf{q}=g_{1,\mathbf{q}}-g_{2,\mathbf{q}}$ for real $g_{j,\mathbf{q}}$. These correlation functions become one when the coupling strength to the common mode are the same.
This reflects the fact that the common mode itself is underdamped.

\section{LAPLACE TRANSFORMATION OF THE INTEGRAL EQUATIONS}

The time-convolution integral in Eqs.~(\ref{eq_dim_integ_eq}) seriously prevents us from obtaining an analytical solution. An alternative way is to invoke the Laplace transformation,
defined as $\mathcal{L}\left\{f(t)\right\}=\int_0^\infty e^{-st}f(t)dt$, transforming the coupled integral equations into a set of linear algebraic equations with the prescriptions
$\mathcal{L}\left\{\int_0^\infty f(t-\tau)g(\tau)d\tau\right\}=\mathcal{L}\left\{f\right\}\cdot\mathcal{L}\left\{g\right\}$ and $\mathcal{L}\left\{\int_0^\infty f(\tau)d\tau\right\}=\mathcal{L}\left\{f\right\}/s$. The transformed equations can be formally expressed as a linear equation:
\begin{equation}
\mathcal{M}\cdot\vec{\rho}_\mathrm{RC}^\mathcal{L}(s)=\vec{\rho}_\mathrm{RC}(0), \label{eq_lap_tran_linear_eq}
\end{equation}
where $\vec{\rho}_\mathrm{RC}^\mathcal{L}(s)$ are the Laplace transformed RC density matrix elements aligned into a column vector. Then, the usual Cramer's rule is helpful for solving
the linear equation.

After careful analysis of the residue properties, we acquire that the steady-state solution in time space is in fact the residue at $s=0$. This observation greatly reduces the necessary
effort in our work.

\section{LAPLACE TRANSFORMATION OF THE CORRELATION FUNCTIONS}

A key element in solving Eq.~(\ref{eq_lap_tran_linear_eq}) consists of the Laplace transformation of the correlation functions. With the help of Bessel's generating function:
\begin{equation}
e^{\frac{x}{2}(t-t^{-1})}=\sum_{n=-\infty}^\infty J_n(x)t^n.
\end{equation}
Here we explicitly write down the results:
\begin{eqnarray}
\mathcal{L}\left\{C(t)\right\}(s)&=&e^{-\frac{|\Delta g_\mathbf{q}|^2}{\omega_\mathbf{q}^2}\coth\left(\frac{\hbar\omega_\mathbf{q}}{2k_\mathrm{B}\mathrm{T}}\right)}
\sum_{n=-\infty}^\infty\sum_{m=0}^\infty\frac{J_n\left(-\frac{|\Delta g_\mathbf{q}|^2}{\omega_\mathbf{q}^2}\right)
\left(\frac{|\Delta g_\mathbf{q}|^2}{\omega_\mathbf{q}^2}\coth\left(\frac{\hbar\omega_\mathbf{q}}{2k_\mathrm{B}\mathrm{T}_\mathrm{a}}\right)\right)^m}{m!}  \nonumber\\
&&\times\sum_{j=0}^{\lfloor\frac{m}{2}\rfloor}\frac{\frac{m!}{(m-2j)!}\omega_{\mathbf{q}}^{2j}(s-in\omega_{\mathbf{q}})}
{\prod_{k=0}^{j}\left((s-in\omega_{\mathbf{q}})^{2}+(m-2k)^{2}\omega_{\mathbf{q}}^{2}\right)},                                                             \\
\mathcal{L}\left\{C^\prime(t)\right\}(s)&=&e^{-\frac{|\Delta g_\mathbf{q}|^2}{\omega_\mathbf{q}^2}\coth\left(\frac{\hbar\omega_\mathbf{q}}{2k_\mathrm{B}\mathrm{T}_\mathrm{a}}\right)}
\sum_{n=-\infty}^\infty\sum_{m=0}^\infty\frac{J_n\left(\frac{|g_{1,\mathbf{q}}|^2-|g_{2,\mathbf{q}}|^2}{\omega_\mathbf{q}^2}\right)
\left(\frac{|\Delta g_\mathbf{q}|^2}{\omega_\mathbf{q}^2}\coth\left(\frac{\hbar\omega_\mathbf{q}}{2k_\mathrm{B}\mathrm{T}}\right)\right)^m}{m!}  \nonumber\\
&&\times\sum_{j=0}^{\lfloor\frac{m}{2}\rfloor}\frac{\frac{m!}{(m-2j)!}\omega_{\mathbf{q}}^{2j}(s-in\omega_{\mathbf{q}})}
{\prod_{k=0}^{j}\left((s-in\omega_{\mathbf{q}})^{2}+(m-2k)^{2}\omega_{\mathbf{q}}^{2}\right)},
\end{eqnarray}
with $\Delta g_\mathbf{q}=g_{1,\mathbf{q}}-g_{2,\mathbf{q}}$ for real $g_{j,\mathbf{q}}$.
\end{widetext}


\begin{thebibliography}{43}%
\makeatletter
\providecommand \@ifxundefined [1]{%
 \@ifx{#1\undefined}
}%
\providecommand \@ifnum [1]{%
 \ifnum #1\expandafter \@firstoftwo
 \else \expandafter \@secondoftwo
 \fi
}%
\providecommand \@ifx [1]{%
 \ifx #1\expandafter \@firstoftwo
 \else \expandafter \@secondoftwo
 \fi
}%
\providecommand \natexlab [1]{#1}%
\providecommand \enquote  [1]{``#1''}%
\providecommand \bibnamefont  [1]{#1}%
\providecommand \bibfnamefont [1]{#1}%
\providecommand \citenamefont [1]{#1}%
\providecommand \href@noop [0]{\@secondoftwo}%
\providecommand \href [0]{\begingroup \@sanitize@url \@href}%
\providecommand \@href[1]{\@@startlink{#1}\@@href}%
\providecommand \@@href[1]{\endgroup#1\@@endlink}%
\providecommand \@sanitize@url [0]{\catcode `\\12\catcode `\$12\catcode
  `\&12\catcode `\#12\catcode `\^12\catcode `\_12\catcode `\%12\relax}%
\providecommand \@@startlink[1]{}%
\providecommand \@@endlink[0]{}%
\providecommand \url  [0]{\begingroup\@sanitize@url \@url }%
\providecommand \@url [1]{\endgroup\@href {#1}{\urlprefix }}%
\providecommand \urlprefix  [0]{URL }%
\providecommand \Eprint [0]{\href }%
\providecommand \doibase [0]{http://dx.doi.org/}%
\providecommand \selectlanguage [0]{\@gobble}%
\providecommand \bibinfo  [0]{\@secondoftwo}%
\providecommand \bibfield  [0]{\@secondoftwo}%
\providecommand \translation [1]{[#1]}%
\providecommand \BibitemOpen [0]{}%
\providecommand \bibitemStop [0]{}%
\providecommand \bibitemNoStop [0]{.\EOS\space}%
\providecommand \EOS [0]{\spacefactor3000\relax}%
\providecommand \BibitemShut  [1]{\csname bibitem#1\endcsname}%
\let\auto@bib@innerbib\@empty
\bibitem [{\citenamefont {Blankenship}(2002)}]{blankenship_text_book}%
  \BibitemOpen
  \bibfield  {author} {\bibinfo {author} {\bibfnamefont {R.~E.}\ \bibnamefont
  {Blankenship}},\ }\href@noop {} {\emph {\bibinfo {title} {Molecular
  Mechanisms of Photosynthesis}}}\ (\bibinfo  {publisher} {Blackwell Science
  Ltd},\ \bibinfo {address} {Oxford, UK},\ \bibinfo {year} {2002})\BibitemShut
  {NoStop}%
\bibitem [{\citenamefont {van Amerongen}\ \emph {et~al.}(2000)\citenamefont
  {van Amerongen}, \citenamefont {Valkunas},\ and\ \citenamefont {van
  Grondelle}}]{grondelle_text_book}%
  \BibitemOpen
  \bibfield  {author} {\bibinfo {author} {\bibfnamefont {H.}~\bibnamefont {van
  Amerongen}}, \bibinfo {author} {\bibfnamefont {L.}~\bibnamefont {Valkunas}},
  \ and\ \bibinfo {author} {\bibfnamefont {R.}~\bibnamefont {van Grondelle}},\
  }\href@noop {} {\emph {\bibinfo {title} {Photosynthetic excitons}}}\
  (\bibinfo  {publisher} {World Scientific Pub Co Inc},\ \bibinfo {address}
  {Singapore},\ \bibinfo {year} {2000})\BibitemShut {NoStop}%
\bibitem [{\citenamefont {Lambert}\ \emph {et~al.}(2013)\citenamefont
  {Lambert}, \citenamefont {Chen}, \citenamefont {Cheng}, \citenamefont {Li},
  \citenamefont {Chen},\ and\ \citenamefont
  {Nori}}]{neill_review_nat_phys_2013}%
  \BibitemOpen
  \bibfield  {author} {\bibinfo {author} {\bibfnamefont {N.}~\bibnamefont
  {Lambert}}, \bibinfo {author} {\bibfnamefont {Y.-N.}\ \bibnamefont {Chen}},
  \bibinfo {author} {\bibfnamefont {Y.-C.}\ \bibnamefont {Cheng}}, \bibinfo
  {author} {\bibfnamefont {C.-M.}\ \bibnamefont {Li}}, \bibinfo {author}
  {\bibfnamefont {G.-Y.}\ \bibnamefont {Chen}}, \ and\ \bibinfo {author}
  {\bibfnamefont {F.}~\bibnamefont {Nori}},\ }\bibinfo{title}{Quantum biology}, \href {\doibase
  10.1038/nphys2474} {\bibfield  {journal} {\bibinfo  {journal} {Nat. Phys.}\
  }\textbf {\bibinfo {volume} {9}},\ \bibinfo {pages} {10} (\bibinfo {year}
  {2013})}\BibitemShut {NoStop}%
\bibitem [{\citenamefont {Lee}\ \emph {et~al.}(2007)\citenamefont {Lee},
  \citenamefont {Cheng},\ and\ \citenamefont {Fleming}}]{fleming_science_2007}%
  \BibitemOpen
  \bibfield  {author} {\bibinfo {author} {\bibfnamefont {H.}~\bibnamefont
  {Lee}}, \bibinfo {author} {\bibfnamefont {Y.-C.}\ \bibnamefont {Cheng}}, \
  and\ \bibinfo {author} {\bibfnamefont {G.~R.}\ \bibnamefont {Fleming}},\
  }\bibinfo{title}{Coherence Dynamics in Photosynthesis: Protein Protection of Excitonic Coherence},
  \href {\doibase 10.1126/science.1142188} {\bibfield  {journal} {\bibinfo
  {journal} {Science}\ }\textbf {\bibinfo {volume} {316}},\ \bibinfo {pages}
  {1462} (\bibinfo {year} {2007})}\BibitemShut {NoStop}%
\bibitem [{\citenamefont {Engel}\ \emph {et~al.}(2007)\citenamefont {Engel},
  \citenamefont {Calhoun}, \citenamefont {Read}, \citenamefont {Ahn},
  \citenamefont {Man\v{c}al}, \citenamefont {Cheng}, \citenamefont
  {Blankenship},\ and\ \citenamefont {Fleming}}]{fleming_nature_2007}%
  \BibitemOpen
  \bibfield  {author} {\bibinfo {author} {\bibfnamefont {G.~S.}\ \bibnamefont
  {Engel}}, \bibinfo {author} {\bibfnamefont {T.~R.}\ \bibnamefont {Calhoun}},
  \bibinfo {author} {\bibfnamefont {E.~L.}\ \bibnamefont {Read}}, \bibinfo
  {author} {\bibfnamefont {T.-K.}\ \bibnamefont {Ahn}}, \bibinfo {author}
  {\bibfnamefont {T.}~\bibnamefont {Man\v{c}al}}, \bibinfo {author}
  {\bibfnamefont {Y.-C.}\ \bibnamefont {Cheng}}, \bibinfo {author}
  {\bibfnamefont {R.~E.}\ \bibnamefont {Blankenship}}, \ and\ \bibinfo {author}
  {\bibfnamefont {G.~R.}\ \bibnamefont {Fleming}},\ }\bibinfo{title}{Evidence for wavelike energy transfer through quantum coherence in photosynthetic systems}, \href {\doibase
  10.1038/nature05678} {\bibfield  {journal} {\bibinfo  {journal} {Nature}\
  }\textbf {\bibinfo {volume} {446}},\ \bibinfo {pages} {782} (\bibinfo {year}
  {2007})}\BibitemShut {NoStop}%
\bibitem [{\citenamefont {Panitchayangkoon}\ \emph {et~al.}(2010)\citenamefont
  {Panitchayangkoon}, \citenamefont {Hayes}, \citenamefont {Fransted},
  \citenamefont {Caram}, \citenamefont {Harel}, \citenamefont {Wen},
  \citenamefont {Blankenship},\ and\ \citenamefont
  {Engel}}]{beating_signals_pnas_2010}%
  \BibitemOpen
  \bibfield  {author} {\bibinfo {author} {\bibfnamefont {G.}~\bibnamefont
  {Panitchayangkoon}}, \bibinfo {author} {\bibfnamefont {D.}~\bibnamefont
  {Hayes}}, \bibinfo {author} {\bibfnamefont {K.~A.}\ \bibnamefont {Fransted}},
  \bibinfo {author} {\bibfnamefont {J.~R.}\ \bibnamefont {Caram}}, \bibinfo
  {author} {\bibfnamefont {E.}~\bibnamefont {Harel}}, \bibinfo {author}
  {\bibfnamefont {J.}~\bibnamefont {Wen}}, \bibinfo {author} {\bibfnamefont
  {R.~E.}\ \bibnamefont {Blankenship}}, \ and\ \bibinfo {author} {\bibfnamefont
  {G.~S.}\ \bibnamefont {Engel}},\ }\bibinfo{title}{Long-lived quantum coherence in photosynthetic complexes at physiological temperature}, \href {\doibase 10.1073/pnas.1005484107}
  {\bibfield  {journal} {\bibinfo  {journal} {Proc. Natl. Acad. Sci. U.S.A.}\
  }\textbf {\bibinfo {volume} {107}},\ \bibinfo {pages} {12766} (\bibinfo
  {year} {2010})}\BibitemShut {NoStop}%
\bibitem [{\citenamefont {Plenio}\ and\ \citenamefont
  {Huelga}(2008)}]{plenio_noise_transport_njp_2008}%
  \BibitemOpen
  \bibfield  {author} {\bibinfo {author} {\bibfnamefont {M.~B.}\ \bibnamefont
  {Plenio}}\ and\ \bibinfo {author} {\bibfnamefont {S.~F.}\ \bibnamefont
  {Huelga}},\ }\bibinfo{title}{Dephasing-assisted transport: quantum networks and biomolecules}, \href {http://stacks.iop.org/1367-2630/10/i=11/a=113019}
  {\bibfield  {journal} {\bibinfo  {journal} {New J. Phys.}\ }\textbf {\bibinfo
  {volume} {10}},\ \bibinfo {pages} {113019} (\bibinfo {year}
  {2008})}\BibitemShut {NoStop}%
\bibitem [{\citenamefont {Caruso}\ \emph {et~al.}(2009)\citenamefont {Caruso},
  \citenamefont {Chin}, \citenamefont {Datta}, \citenamefont {Huelga},\ and\
  \citenamefont {Plenio}}]{plenio_noise_transport_jcp_2009}%
  \BibitemOpen
  \bibfield  {author} {\bibinfo {author} {\bibfnamefont {F.}~\bibnamefont
  {Caruso}}, \bibinfo {author} {\bibfnamefont {A.~W.}\ \bibnamefont {Chin}},
  \bibinfo {author} {\bibfnamefont {A.}~\bibnamefont {Datta}}, \bibinfo
  {author} {\bibfnamefont {S.~F.}\ \bibnamefont {Huelga}}, \ and\ \bibinfo
  {author} {\bibfnamefont {M.~B.}\ \bibnamefont {Plenio}},\ }\bibinfo{title}{Highly efficient energy excitation transfer in light-harvesting complexes: The fundamental role of noise-assisted transport}, \href {\doibase
  10.1063/1.3223548} {\bibfield  {journal} {\bibinfo
  {journal} {J. Chem. Phys.}\ }\textbf {\bibinfo {volume} {131}},\ \bibinfo
  {pages} {105106} (\bibinfo {year} {2009})}\BibitemShut {NoStop}%
\bibitem [{\citenamefont {Romero}\ \emph {et~al.}(2014)\citenamefont {Romero},
  \citenamefont {Augulis}, \citenamefont {Novoderezhkin}, \citenamefont
  {Ferretti}, \citenamefont {Thieme}, \citenamefont {Zigmantas},\ and\
  \citenamefont {van Grondelle}}]{elisabet_quan_coh_nat_phys_2014}%
  \BibitemOpen
  \bibfield  {author} {\bibinfo {author} {\bibfnamefont {E.}~\bibnamefont
  {Romero}}, \bibinfo {author} {\bibfnamefont {R.}~\bibnamefont {Augulis}},
  \bibinfo {author} {\bibfnamefont {V.~I.}\ \bibnamefont {Novoderezhkin}},
  \bibinfo {author} {\bibfnamefont {M.}~\bibnamefont {Ferretti}}, \bibinfo
  {author} {\bibfnamefont {J.}~\bibnamefont {Thieme}}, \bibinfo {author}
  {\bibfnamefont {D.}~\bibnamefont {Zigmantas}}, \ and\ \bibinfo {author}
  {\bibfnamefont {R.}~\bibnamefont {van Grondelle}},\ }\bibinfo{title}{Quantum coherence in photosynthesis for efficient solar-energy conversion}, \href {\doibase
  10.1038/nphys3017} {\bibfield  {journal} {\bibinfo  {journal} {Nat. Phys.}\
  }\textbf {\bibinfo {volume} {10}},\ \bibinfo {pages} {676} (\bibinfo {year}
  {2014})}\BibitemShut {NoStop}%
\bibitem [{\citenamefont {Pach\'{o}n}\ and\ \citenamefont
  {Brumer}(2011)}]{pachon_origin_coherence_jpcl_2011}%
  \BibitemOpen
  \bibfield  {author} {\bibinfo {author} {\bibfnamefont {L.~A.}\ \bibnamefont
  {Pach\'{o}n}}\ and\ \bibinfo {author} {\bibfnamefont {P.}~\bibnamefont
  {Brumer}},\ }\bibinfo{title}{Physical Basis for Long-Lived Electronic Coherence in Photosynthetic Light-Harvesting Systems}, \href {\doibase 10.1021/jz201189p} {\bibfield  {journal}
  {\bibinfo  {journal} {J. Phys. Chem. Lett.}\ }\textbf {\bibinfo {volume}
  {2}},\ \bibinfo {pages} {2728} (\bibinfo {year} {2011})}\BibitemShut
  {NoStop}%
\bibitem [{\citenamefont {Christensson}\ \emph {et~al.}(2012)\citenamefont
  {Christensson}, \citenamefont {Kauffmann}, \citenamefont {Pullerits},\ and\
  \citenamefont {Man\v{c}al}}]{christensson_origin_coherence_jpcb_2012}%
  \BibitemOpen
  \bibfield  {author} {\bibinfo {author} {\bibfnamefont {N.}~\bibnamefont
  {Christensson}}, \bibinfo {author} {\bibfnamefont {H.~F.}\ \bibnamefont
  {Kauffmann}}, \bibinfo {author} {\bibfnamefont {T.}~\bibnamefont
  {Pullerits}}, \ and\ \bibinfo {author} {\bibfnamefont {T.}~\bibnamefont
  {Man\v{c}al}},\ } \bibinfo{title}{Origin of Long-Lived Coherences in Light-Harvesting Complexes}, \href {\doibase 10.1021/jp304649c} {\bibfield  {journal}
  {\bibinfo  {journal} {J. Phys. Chem. B}\ }\textbf {\bibinfo {volume} {116}},\
  \bibinfo {pages} {7449} (\bibinfo {year} {2012})}\BibitemShut {NoStop}%
\bibitem [{\citenamefont {Plenio}\ \emph {et~al.}(2013)\citenamefont {Plenio},
  \citenamefont {Almeida},\ and\ \citenamefont
  {Huelga}}]{plenio_long_coh_jcp_2013}%
  \BibitemOpen
  \bibfield  {author} {\bibinfo {author} {\bibfnamefont {M.~B.}\ \bibnamefont
  {Plenio}}, \bibinfo {author} {\bibfnamefont {J.}~\bibnamefont {Almeida}}, \
  and\ \bibinfo {author} {\bibfnamefont {S.~F.}\ \bibnamefont {Huelga}},\
  }\bibinfo{title}{Origin of long-lived oscillations in 2D-spectra of a quantum vibronic model: Electronic versus vibrational coherence}, \href {\doibase 10.1063/1.4846275} {\bibfield  {journal} {\bibinfo
  {journal} {J. Chem. Phys.}\ }\textbf {\bibinfo {volume} {139}},\ \bibinfo
  {pages} {235102} (\bibinfo {year} {2013})}\BibitemShut {NoStop}%
\bibitem [{\citenamefont {Caram}\ \emph {et~al.}(2012)\citenamefont {Caram},
  \citenamefont {Lewis}, \citenamefont {Fidler},\ and\ \citenamefont
  {Engel}}]{engel_corr_fluc_jcp_2012}%
  \BibitemOpen
  \bibfield  {author} {\bibinfo {author} {\bibfnamefont {J.~R.}\ \bibnamefont
  {Caram}}, \bibinfo {author} {\bibfnamefont {N.~H.~C.}\ \bibnamefont {Lewis}},
  \bibinfo {author} {\bibfnamefont {A.~F.}\ \bibnamefont {Fidler}}, \ and\
  \bibinfo {author} {\bibfnamefont {G.~S.}\ \bibnamefont {Engel}},\ }\bibinfo{title}{Signatures of correlated excitonic dynamics in two-dimensional spectroscopy of the Fenna-Matthew-Olson photosynthetic complex}, \href
  {\doibase 10.1063/1.3690498} {\bibfield  {journal} {\bibinfo  {journal} {J.
  Chem. Phys.}\ }\textbf {\bibinfo {volume} {136}},\ \bibinfo {pages} {104505}
  (\bibinfo {year} {2012})}\BibitemShut {NoStop}%
\bibitem [{\citenamefont {Kolli}\ \emph {et~al.}(2012)\citenamefont {Kolli},
  \citenamefont {O'Reilly}, \citenamefont {Scholes},\ and\ \citenamefont
  {Olaya-Castro}}]{avinash_jcp_2012}%
  \BibitemOpen
  \bibfield  {author} {\bibinfo {author} {\bibfnamefont {A.}~\bibnamefont
  {Kolli}}, \bibinfo {author} {\bibfnamefont {E.~J.}\ \bibnamefont
  {O'Reilly}}, \bibinfo {author} {\bibfnamefont {G.~D.}\ \bibnamefont
  {Scholes}}, \ and\ \bibinfo {author} {\bibfnamefont {A.}~\bibnamefont
  {Olaya-Castro}},\ }\bibinfo{title}{The fundamental role of quantized vibrations in coherent light harvesting by cryptophyte algae}, \href {\doibase 10.1063/1.4764100} {\bibfield  {journal}
  {\bibinfo  {journal} {J. Chem. Phys.}\ }\textbf {\bibinfo {volume} {137}},\
  \bibinfo {pages} {174109} (\bibinfo {year} {2012})}\BibitemShut {NoStop}%
\bibitem [{\citenamefont {Chin}\ \emph {et~al.}(2013)\citenamefont {Chin},
  \citenamefont {Prior}, \citenamefont {Rosenbach}, \citenamefont
  {Caycedo-Soler}, \citenamefont {Huelga},\ and\ \citenamefont
  {Plenio}}]{plenio_nature_physics_2013}%
  \BibitemOpen
  \bibfield  {author} {\bibinfo {author} {\bibfnamefont {A.~W.}\ \bibnamefont
  {Chin}}, \bibinfo {author} {\bibfnamefont {J.}~\bibnamefont {Prior}},
  \bibinfo {author} {\bibfnamefont {R.}~\bibnamefont {Rosenbach}}, \bibinfo
  {author} {\bibfnamefont {F.}~\bibnamefont {Caycedo-Soler}}, \bibinfo {author}
  {\bibfnamefont {S.~F.}\ \bibnamefont {Huelga}}, \ and\ \bibinfo {author}
  {\bibfnamefont {M.~B.}\ \bibnamefont {Plenio}},\ }\bibinfo{title}{The role of non-equilibrium vibrational structures in electronic coherence and recoherence in pigment-protein complexes}, \href {\doibase
  10.1038/nphys2515} {\bibfield  {journal} {\bibinfo  {journal} {Nat. Phys.}\
  }\textbf {\bibinfo {volume} {9}},\ \bibinfo {pages} {113} (\bibinfo {year}
  {2013})}\BibitemShut {NoStop}%
\bibitem [{\citenamefont {Chen}\ \emph {et~al.}(2014)\citenamefont {Chen},
  \citenamefont {Lien}, \citenamefont {Hwang},\ and\ \citenamefont
  {Chen}}]{hongbin_pre_2014}%
  \BibitemOpen
  \bibfield  {author} {\bibinfo {author} {\bibfnamefont {H.-B.}\ \bibnamefont
  {Chen}}, \bibinfo {author} {\bibfnamefont {J.-Y.}\ \bibnamefont {Lien}},
  \bibinfo {author} {\bibfnamefont {C.-C.}\ \bibnamefont {Hwang}}, \ and\
  \bibinfo {author} {\bibfnamefont {Y.-N.}\ \bibnamefont {Chen}},\ }\bibinfo{title}{Long-lived quantum coherence and non-Markovianity of photosynthetic complexes}, \href
  {\doibase 10.1103/PhysRevE.89.042147} {\bibfield  {journal} {\bibinfo
  {journal} {Phys. Rev. E}\ }\textbf {\bibinfo {volume} {89}},\ \bibinfo
  {pages} {042147} (\bibinfo {year} {2014})}\BibitemShut {NoStop}%
\bibitem [{\citenamefont {Chen}\ \emph {et~al.}(2015)\citenamefont {Chen},
  \citenamefont {Lambert}, \citenamefont {Cheng}, \citenamefont {Chen},\ and\
  \citenamefont {Nori}}]{hongbin_scirep_2015}%
  \BibitemOpen
  \bibfield  {author} {\bibinfo {author} {\bibfnamefont {H.-B.}\ \bibnamefont
  {Chen}}, \bibinfo {author} {\bibfnamefont {N.}~\bibnamefont {Lambert}},
  \bibinfo {author} {\bibfnamefont {Y.-C.}\ \bibnamefont {Cheng}}, \bibinfo
  {author} {\bibfnamefont {Y.-N.}\ \bibnamefont {Chen}}, \ and\ \bibinfo
  {author} {\bibfnamefont {F.}~\bibnamefont {Nori}},\ }\bibinfo{title}{Using non-Markovian measures to evaluate quantum master equations for photosynthesis}, \href {\doibase
  10.1038/srep12753} {\bibfield  {journal} {\bibinfo  {journal} {Sci. Rep.}\
  }\textbf {\bibinfo {volume} {5}},\ \bibinfo {pages} {12753} (\bibinfo {year}
  {2015})}\BibitemShut {NoStop}%
\bibitem [{\citenamefont {Dorfman}\ \emph {et~al.}(2013)\citenamefont
  {Dorfman}, \citenamefont {Voronine}, \citenamefont {Mukamel},\ and\
  \citenamefont {Scully}}]{dorfman_bio_qhe_pnas_2013}%
  \BibitemOpen
  \bibfield  {author} {\bibinfo {author} {\bibfnamefont {K.~E.}\ \bibnamefont
  {Dorfman}}, \bibinfo {author} {\bibfnamefont {D.~V.}\ \bibnamefont
  {Voronine}}, \bibinfo {author} {\bibfnamefont {S.}~\bibnamefont {Mukamel}}, \
  and\ \bibinfo {author} {\bibfnamefont {M.~O.}\ \bibnamefont {Scully}},\
  }\bibinfo{title}{Photosynthetic reaction center as a quantum heat engine}, \href {\doibase 10.1073/pnas.1212666110} {\bibfield  {journal} {\bibinfo
  {journal} {Proc. Natl. Acad. Sci. U.S.A.}\ }\textbf {\bibinfo {volume}
  {110}},\ \bibinfo {pages} {2746} (\bibinfo {year} {2013})}\BibitemShut
  {NoStop}%
\bibitem [{\citenamefont {Creatore}\ \emph {et~al.}(2013)\citenamefont
  {Creatore}, \citenamefont {Parker}, \citenamefont {Emmott},\ and\
  \citenamefont {Chin}}]{creatore_bio_qhe_prl_2013}%
  \BibitemOpen
  \bibfield  {author} {\bibinfo {author} {\bibfnamefont {C.}~\bibnamefont
  {Creatore}}, \bibinfo {author} {\bibfnamefont {M.~A.}\ \bibnamefont
  {Parker}}, \bibinfo {author} {\bibfnamefont {S.}~\bibnamefont {Emmott}}, \
  and\ \bibinfo {author} {\bibfnamefont {A.~W.}\ \bibnamefont {Chin}},\ }\bibinfo{title}{Efficient Biologically Inspired Photocell Enhanced by Delocalized Quantum States}, \href
  {\doibase 10.1103/PhysRevLett.111.253601} {\bibfield  {journal} {\bibinfo
  {journal} {Phys. Rev. Lett.}\ }\textbf {\bibinfo {volume} {111}},\ \bibinfo
  {pages} {253601} (\bibinfo {year} {2013})}\BibitemShut {NoStop}%
\bibitem [{\citenamefont {Einstein}(1917)}]{einstein_phys_z_1917}%
  \BibitemOpen
  \bibfield  {author} {\bibinfo {author} {\bibfnamefont {A.}~\bibnamefont
  {Einstein}},\ }\bibinfo{title}{Zur Quantentheorie der Strahlung}, \href@noop {} {\bibfield  {journal} {\bibinfo  {journal}
  {Phys. Z.}\ }\textbf {\bibinfo {volume} {18}},\ \bibinfo {pages} {121}
  (\bibinfo {year} {1917})}\BibitemShut {NoStop}%
\bibitem [{\citenamefont {Scovil}\ and\ \citenamefont
  {Schulz-DuBois}(1959)}]{maser_effi_prl_1959}%
  \BibitemOpen
  \bibfield  {author} {\bibinfo {author} {\bibfnamefont {H.~E.~D.}\
  \bibnamefont {Scovil}}\ and\ \bibinfo {author} {\bibfnamefont {E.~O.}\
  \bibnamefont {Schulz-DuBois}},\ }\bibinfo{title}{Three-Level Masers as Heat Engines}, \href {\doibase 10.1103/PhysRevLett.2.262}
  {\bibfield  {journal} {\bibinfo  {journal} {Phys. Rev. Lett.}\ }\textbf
  {\bibinfo {volume} {2}},\ \bibinfo {pages} {262} (\bibinfo {year}
  {1959})}\BibitemShut {NoStop}%
\bibitem [{\citenamefont {Shockley}\ and\ \citenamefont
  {Queisser}(1961)}]{photocell_effi_jap_1961}%
  \BibitemOpen
  \bibfield  {author} {\bibinfo {author} {\bibfnamefont {W.}~\bibnamefont
  {Shockley}}\ and\ \bibinfo {author} {\bibfnamefont {H.~J.}\ \bibnamefont
  {Queisser}},\ }\bibinfo{title}{Detailed Balance Limit of Efficiency of p-n Junction Solar Cells}, \href {\doibase 10.1063/1.1736034} {\bibfield  {journal}
  {\bibinfo  {journal} {J. Appl. Phys.}\ }\textbf {\bibinfo {volume} {32}},\
  \bibinfo {pages} {510} (\bibinfo {year} {1961})}\BibitemShut {NoStop}%
\bibitem [{\citenamefont {Scully}(2010)}]{scully_qphotocell_prl_2010}%
  \BibitemOpen
  \bibfield  {author} {\bibinfo {author} {\bibfnamefont {M.~O.}\ \bibnamefont
  {Scully}},\ }\bibinfo{title}{Quantum Photocell: Using Quantum Coherence to Reduce Radiative Recombination and Increase Efficiency}, \href {\doibase 10.1103/PhysRevLett.104.207701} {\bibfield
  {journal} {\bibinfo  {journal} {Phys. Rev. Lett.}\ }\textbf {\bibinfo
  {volume} {104}},\ \bibinfo {pages} {207701} (\bibinfo {year}
  {2010})}\BibitemShut {NoStop}%
\bibitem [{\citenamefont {Scully}\ \emph {et~al.}(2011)\citenamefont {Scully},
  \citenamefont {Chapin}, \citenamefont {Dorfman}, \citenamefont {Kim},\ and\
  \citenamefont {Svidzinsky}}]{scully_qhe_pnas_2011}%
  \BibitemOpen
  \bibfield  {author} {\bibinfo {author} {\bibfnamefont {M.~O.}\ \bibnamefont
  {Scully}}, \bibinfo {author} {\bibfnamefont {K.~R.}\ \bibnamefont {Chapin}},
  \bibinfo {author} {\bibfnamefont {K.~E.}\ \bibnamefont {Dorfman}}, \bibinfo
  {author} {\bibfnamefont {M.~B.}\ \bibnamefont {Kim}}, \ and\ \bibinfo
  {author} {\bibfnamefont {A.}~\bibnamefont {Svidzinsky}},\ }\bibinfo{title}{Quantum heat engine power can be increased by noise-induced coherence}, \href {\doibase
  10.1073/pnas.1110234108} {\bibfield  {journal} {\bibinfo  {journal} {Proc.
  Natl. Acad. Sci. U.S.A.}\ }\textbf {\bibinfo {volume} {108}},\ \bibinfo
  {pages} {15097} (\bibinfo {year} {2011})}\BibitemShut {NoStop}%
\bibitem [{\citenamefont {Svidzinsky}\ \emph {et~al.}(2011)\citenamefont
  {Svidzinsky}, \citenamefont {Dorfman},\ and\ \citenamefont
  {Scully}}]{svidzinsky_fano_int_pra_2011}%
  \BibitemOpen
  \bibfield  {author} {\bibinfo {author} {\bibfnamefont {A.~A.}\ \bibnamefont
  {Svidzinsky}}, \bibinfo {author} {\bibfnamefont {K.~E.}\ \bibnamefont
  {Dorfman}}, \ and\ \bibinfo {author} {\bibfnamefont {M.~O.}\ \bibnamefont
  {Scully}},\ }\bibinfo{title}{Enhancing photovoltaic power by Fano-induced coherence}, \href {\doibase 10.1103/PhysRevA.84.053818} {\bibfield
  {journal} {\bibinfo  {journal} {Phys. Rev. A}\ }\textbf {\bibinfo {volume}
  {84}},\ \bibinfo {pages} {053818} (\bibinfo {year} {2011})}\BibitemShut
  {NoStop}%
\bibitem [{\citenamefont {Zhang}\ \emph {et~al.}(2015)\citenamefont {Zhang},
  \citenamefont {Oh}, \citenamefont {Alharbi}, \citenamefont {Engel},\ and\
  \citenamefont {Kais}}]{yiteng_pccp_2015}%
  \BibitemOpen
  \bibfield  {author} {\bibinfo {author} {\bibfnamefont {Y.}~\bibnamefont
  {Zhang}}, \bibinfo {author} {\bibfnamefont {S.}~\bibnamefont {Oh}}, \bibinfo
  {author} {\bibfnamefont {F.~H.}\ \bibnamefont {Alharbi}}, \bibinfo {author}
  {\bibfnamefont {G.~S.}\ \bibnamefont {Engel}}, \ and\ \bibinfo {author}
  {\bibfnamefont {S.}~\bibnamefont {Kais}},\ }\bibinfo{title}{Delocalized quantum states enhance photocell efficiency}, \href {\doibase
  10.1039/C4CP05310A} {\bibfield  {journal} {\bibinfo  {journal} {Phys. Chem.
  Chem. Phys.}\ }\textbf {\bibinfo {volume} {17}},\ \bibinfo {pages} {5743}
  (\bibinfo {year} {2015})}\BibitemShut {NoStop}%
\bibitem [{\citenamefont {Peterman}\ \emph {et~al.}(1997)\citenamefont
  {Peterman}, \citenamefont {Pullerits}, \citenamefont {van Grondelle},\ and\
  \citenamefont {van Amerongen}}]{grondelle_vib_phonon_jpcb_1997}%
  \BibitemOpen
  \bibfield  {author} {\bibinfo {author} {\bibfnamefont {E.~J.~G.}\
  \bibnamefont {Peterman}}, \bibinfo {author} {\bibfnamefont {T.}~\bibnamefont
  {Pullerits}}, \bibinfo {author} {\bibfnamefont {R.}~\bibnamefont {van
  Grondelle}}, \ and\ \bibinfo {author} {\bibfnamefont {H.}~\bibnamefont {van
  Amerongen}},\ }\bibinfo{title}{Electron-Phonon Coupling and Vibronic Fine Structure of Light-Harvesting Complex II of Green Plants:  Temperature Dependent Absorption and High-Resolution Fluorescence Spectroscopy},
  \href {\doibase 10.1021/jp962338e} {\bibfield  {journal}
  {\bibinfo  {journal} {J. Phys. Chem. B}\ }\textbf {\bibinfo {volume} {101}},\
  \bibinfo {pages} {4448} (\bibinfo {year} {1997})}\BibitemShut {NoStop}%
\bibitem [{\citenamefont {Wendling}\ \emph {et~al.}(2000)\citenamefont
  {Wendling}, \citenamefont {Pullerits}, \citenamefont {Przyjalgowski},
  \citenamefont {Vulto}, \citenamefont {Aartsma}, \citenamefont {van
  Grondelle},\ and\ \citenamefont {van
  Amerongen}}]{grondelle_vib_phonon_jpcb_2000}%
  \BibitemOpen
  \bibfield  {author} {\bibinfo {author} {\bibfnamefont {M.}~\bibnamefont
  {Wendling}}, \bibinfo {author} {\bibfnamefont {T.}~\bibnamefont {Pullerits}},
  \bibinfo {author} {\bibfnamefont {M.~A.}\ \bibnamefont {Przyjalgowski}},
  \bibinfo {author} {\bibfnamefont {S.~I.~E.}\ \bibnamefont {Vulto}}, \bibinfo
  {author} {\bibfnamefont {T.~J.}\ \bibnamefont {Aartsma}}, \bibinfo {author}
  {\bibfnamefont {R.}~\bibnamefont {van Grondelle}}, \ and\ \bibinfo {author}
  {\bibfnamefont {H.}~\bibnamefont {van Amerongen}},\ }\bibinfo{title}{Electron-Vibrational Coupling in the Fenna-Matthews-Olson Complex of Prosthecochloris aestuarii Determined by Temperature-Dependent Absorption and Fluorescence Line-Narrowing Measurements}, \href {\doibase
  10.1021/jp000077+} {\bibfield  {journal} {\bibinfo  {journal} {J. Phys. Chem.
  B}\ }\textbf {\bibinfo {volume} {104}},\ \bibinfo {pages} {5825} (\bibinfo
  {year} {2000})}\BibitemShut {NoStop}%
\bibitem [{\citenamefont {Womick}\ and\ \citenamefont
  {Moran}(2011)}]{jordan_vib_enh_eet_jpcb_2011}%
  \BibitemOpen
  \bibfield  {author} {\bibinfo {author} {\bibfnamefont {J.~M.}\ \bibnamefont
  {Womick}}\ and\ \bibinfo {author} {\bibfnamefont {A.~M.}\ \bibnamefont
  {Moran}},\ }\bibinfo{title}{Vibronic Enhancement of Exciton Sizes and Energy Transport in Photosynthetic Complexes}, \href {\doibase 10.1021/jp106713q} {\bibfield  {journal}
  {\bibinfo  {journal} {J. Phys. Chem. B}\ }\textbf {\bibinfo {volume} {115}},\
  \bibinfo {pages} {1347} (\bibinfo {year} {2011})}\BibitemShut {NoStop}%
\bibitem [{\citenamefont {Tiwari}\ \emph {et~al.}(2013)\citenamefont {Tiwari},
  \citenamefont {Peters},\ and\ \citenamefont {Jonas}}]{vivek_pnas_2013}%
  \BibitemOpen
  \bibfield  {author} {\bibinfo {author} {\bibfnamefont {V.}~\bibnamefont
  {Tiwari}}, \bibinfo {author} {\bibfnamefont {W.~K.}\ \bibnamefont {Peters}},
  \ and\ \bibinfo {author} {\bibfnamefont {D.~M.}\ \bibnamefont {Jonas}},\
  }\bibinfo{title}{Electronic resonance with anticorrelated pigment vibrations drives photosynthetic energy transfer outside the adiabatic framework}, \href {\doibase 10.1073/pnas.1211157110} {\bibfield  {journal} {\bibinfo
  {journal} {Proc. Natl. Acad. Sci. U.S.A.}\ }\textbf {\bibinfo {volume}
  {110}},\ \bibinfo {pages} {1203} (\bibinfo {year} {2013})}\BibitemShut
  {NoStop}%
\bibitem [{\citenamefont {O'Reilly}\ and\ \citenamefont
  {Olaya-Castro}(2014)}]{edward_nat_commun_2014}%
  \BibitemOpen
  \bibfield  {author} {\bibinfo {author} {\bibfnamefont {E.~J.}\ \bibnamefont
  {O'Reilly}}\ and\ \bibinfo {author} {\bibfnamefont {A.}~\bibnamefont
  {Olaya-Castro}},\ }\bibinfo{title}{Non-classicality of the molecular vibrations assisting exciton energy transfer at room temperature}, \href {\doibase 10.1038/ncomms4012} {\bibfield  {journal}
  {\bibinfo  {journal} {Nat. Commun.}\ }\textbf {\bibinfo {volume} {5}},\
  \bibinfo {pages} {3012} (\bibinfo {year} {2014})}\BibitemShut {NoStop}%
\bibitem [{\citenamefont {Loll}\ \emph {et~al.}(2005)\citenamefont {Loll},
  \citenamefont {Kern}, \citenamefont {Saenger}, \citenamefont {Zouni},\ and\
  \citenamefont {Biesiadka}}]{bernhard_ps2_org_nature_2005}%
  \BibitemOpen
  \bibfield  {author} {\bibinfo {author} {\bibfnamefont {B.}~\bibnamefont
  {Loll}}, \bibinfo {author} {\bibfnamefont {J.}~\bibnamefont {Kern}}, \bibinfo
  {author} {\bibfnamefont {W.}~\bibnamefont {Saenger}}, \bibinfo {author}
  {\bibfnamefont {A.}~\bibnamefont {Zouni}}, \ and\ \bibinfo {author}
  {\bibfnamefont {J.}~\bibnamefont {Biesiadka}},\ }\bibinfo{title}{Towards complete cofactor arrangement in the 3.0~\text{\AA} resolution structure of photosystem II}, \href {\doibase
  10.1038/nature04224} {\bibfield  {journal} {\bibinfo  {journal} {Nature}\
  }\textbf {\bibinfo {volume} {438}},\ \bibinfo {pages} {1040} (\bibinfo {year}
  {2005})}\BibitemShut {NoStop}%
\bibitem [{\citenamefont {Kern}\ and\ \citenamefont
  {Renger}(2007)}]{jan_ps2_org_photosynth_res_2007}%
  \BibitemOpen
  \bibfield  {author} {\bibinfo {author} {\bibfnamefont {J.}~\bibnamefont
  {Kern}}\ and\ \bibinfo {author} {\bibfnamefont {G.}~\bibnamefont {Renger}},\
  }\bibinfo{title}{Photosystem II: Structure and mechanism of the water:plastoquinone oxidoreductase}, \href {\doibase 10.1007/s11120-007-9201-1} {\bibfield  {journal} {\bibinfo
  {journal} {Photosynth. Res.}\ }\textbf {\bibinfo {volume} {94}},\ \bibinfo
  {pages} {183} (\bibinfo {year} {2007})}\BibitemShut {NoStop}%
\bibitem [{\citenamefont {Muh}\ \emph {et~al.}(2008)\citenamefont {Muh},
  \citenamefont {Renger},\ and\ \citenamefont {Zouni}}]{mueh_ps2_org_ppb_2008}%
  \BibitemOpen
  \bibfield  {author} {\bibinfo {author} {\bibfnamefont {F.}~\bibnamefont
  {Muh}}, \bibinfo {author} {\bibfnamefont {T.}~\bibnamefont {Renger}}, \ and\
  \bibinfo {author} {\bibfnamefont {A.}~\bibnamefont {Zouni}},\ }\bibinfo{title}{Crystal structure of cyanobacterial photosystem II at 3.0~\text{\AA} resolution: A closer look at the antenna system and the small membrane-intrinsic subunits}, \href
  {\doibase 10.1016/j.plaphy.2008.01.003} {\bibfield  {journal} {\bibinfo
  {journal} {Plant Physiol. Biochem.}\ }\textbf {\bibinfo {volume} {46}},\
  \bibinfo {pages} {238} (\bibinfo {year} {2008})}\BibitemShut {NoStop}%
\bibitem [{\citenamefont {Croce}\ and\ \citenamefont {van
  Amerongen}(2011)}]{roberta_ps2_org_jppb_2011}%
  \BibitemOpen
  \bibfield  {author} {\bibinfo {author} {\bibfnamefont {R.}~\bibnamefont
  {Croce}}\ and\ \bibinfo {author} {\bibfnamefont {H.}~\bibnamefont {van
  Amerongen}},\ }\bibinfo{title}{Light-harvesting and structural organization of Photosystem II: From individual complexes to thylakoid membrane}, \href {\doibase 10.1016/j.jphotobiol.2011.02.015} {\bibfield
  {journal} {\bibinfo  {journal} {J. Photochem. Photobiol. B}\ }\textbf
  {\bibinfo {volume} {104}},\ \bibinfo {pages} {142} (\bibinfo {year}
  {2011})}\BibitemShut {NoStop}%
\bibitem [{\citenamefont {Novoderezhkin}\ \emph {et~al.}(2007)\citenamefont
  {Novoderezhkin}, \citenamefont {Dekker},\ and\ \citenamefont {van
  Grondelle}}]{grondelle_ct_pathway_biophys_j_2007}%
  \BibitemOpen
  \bibfield  {author} {\bibinfo {author} {\bibfnamefont {V.~I.}\ \bibnamefont
  {Novoderezhkin}}, \bibinfo {author} {\bibfnamefont {J.~P.}\ \bibnamefont
  {Dekker}}, \ and\ \bibinfo {author} {\bibfnamefont {R.}~\bibnamefont {van
  Grondelle}},\ }\bibinfo{title}{Mixing of exciton and charge-transfer states in Photosystem II reaction centers: modeling of Stark spectra with modified Redfield theory}, \href {\doibase 10.1529/biophysj.106.096867} {\bibfield
  {journal} {\bibinfo  {journal} {Biophys. J.}\ }\textbf {\bibinfo {volume}
  {93}},\ \bibinfo {pages} {1293} (\bibinfo {year} {2007})}\BibitemShut
  {NoStop}%
\bibitem [{\citenamefont {Romero}\ \emph {et~al.}(2010)\citenamefont {Romero},
  \citenamefont {van Stokkum}, \citenamefont {Novoderezhkin}, \citenamefont
  {Dekker},\ and\ \citenamefont {van
  Grondelle}}]{grondelle_ct_pathway_biochem_2010}%
  \BibitemOpen
  \bibfield  {author} {\bibinfo {author} {\bibfnamefont {E.}~\bibnamefont
  {Romero}}, \bibinfo {author} {\bibfnamefont {I.~H.~M.}\ \bibnamefont {van
  Stokkum}}, \bibinfo {author} {\bibfnamefont {V.~I.}\ \bibnamefont
  {Novoderezhkin}}, \bibinfo {author} {\bibfnamefont {J.~P.}\ \bibnamefont
  {Dekker}}, \ and\ \bibinfo {author} {\bibfnamefont {R.}~\bibnamefont {van
  Grondelle}},\ }\bibinfo{title}{Two Different Charge Separation Pathways in Photosystem II}, \href {\doibase 10.1021/bi1003926} {\bibfield  {journal}
  {\bibinfo  {journal} {Biochemistry}\ }\textbf {\bibinfo {volume} {49}},\
  \bibinfo {pages} {4300} (\bibinfo {year} {2010})}\BibitemShut {NoStop}%
\bibitem [{\citenamefont {Cardona}\ \emph {et~al.}(2012)\citenamefont
  {Cardona}, \citenamefont {Sedoud}, \citenamefont {Cox},\ and\ \citenamefont
  {Rutherford}}]{cardona_ps2_review_bba_2012}%
  \BibitemOpen
  \bibfield  {author} {\bibinfo {author} {\bibfnamefont {T.}~\bibnamefont
  {Cardona}}, \bibinfo {author} {\bibfnamefont {A.}~\bibnamefont {Sedoud}},
  \bibinfo {author} {\bibfnamefont {N.}~\bibnamefont {Cox}}, \ and\ \bibinfo
  {author} {\bibfnamefont {A.~W.}\ \bibnamefont {Rutherford}},\ }\bibinfo{title}{Charge separation in Photosystem II: A comparative and evolutionary overview}, \href
  {\doibase 10.1016/j.bbabio.2011.07.012} {\bibfield  {journal} {\bibinfo
  {journal} {Biochim. Biophys. Acta}\ }\textbf {\bibinfo {volume} {1817}},\
  \bibinfo {pages} {26} (\bibinfo {year} {2012})}\BibitemShut {NoStop}%
\bibitem [{\citenamefont {Fujihashi}\ \emph {et~al.}(2015)\citenamefont
  {Fujihashi}, \citenamefont {Fleming},\ and\ \citenamefont
  {Ishizaki}}]{fujihashi_nuclear_vib_jcp_2015}%
  \BibitemOpen
  \bibfield  {author} {\bibinfo {author} {\bibfnamefont {Y.}~\bibnamefont
  {Fujihashi}}, \bibinfo {author} {\bibfnamefont {G.~R.}\ \bibnamefont
  {Fleming}}, \ and\ \bibinfo {author} {\bibfnamefont {A.}~\bibnamefont
  {Ishizaki}},\ }\bibinfo{title}{Impact of environmentally induced fluctuations on quantum mechanically mixed electronic and vibrational pigment states in photosynthetic energy transfer and 2D electronic spectra}, \href {\doibase 10.1063/1.4914302} {\bibfield  {journal}
  {\bibinfo  {journal} {J. Chem. Phys.}\ }\textbf {\bibinfo {volume} {142}},\
  \bibinfo {eid} {212403} (\bibinfo {year} {2015})}\BibitemShut {NoStop}%
\bibitem [{\citenamefont {Killoran}\ \emph {et~al.}(2015)\citenamefont
  {Killoran}, \citenamefont {Huelga},\ and\ \citenamefont
  {Plenio}}]{planio_vib_bqhe_jcp_2015}%
  \BibitemOpen
  \bibfield  {author} {\bibinfo {author} {\bibfnamefont {N.}~\bibnamefont
  {Killoran}}, \bibinfo {author} {\bibfnamefont {S.~F.}\ \bibnamefont
  {Huelga}}, \ and\ \bibinfo {author} {\bibfnamefont {M.~B.}\ \bibnamefont
  {Plenio}},\ }\bibinfo{title}{Enhancing light-harvesting power with coherent vibrational interactions: A quantum heat engine picture}, \href {\doibase 10.1063/1.4932307} {\bibfield  {journal}
  {\bibinfo  {journal} {J. Chem. Phys.}\ }\textbf {\bibinfo {volume} {143}},\
  \bibinfo {eid} {155102} (\bibinfo {year} {2015})}\BibitemShut {NoStop}%
\bibitem [{\citenamefont {Jang}\ \emph {et~al.}(2008)\citenamefont {Jang},
  \citenamefont {Cheng}, \citenamefont {Reichman},\ and\ \citenamefont
  {Eaves}}]{seogjoo_jang_polaron-1}%
  \BibitemOpen
  \bibfield  {author} {\bibinfo {author} {\bibfnamefont {S.}~\bibnamefont
  {Jang}}, \bibinfo {author} {\bibfnamefont {Y.-C.}\ \bibnamefont {Cheng}},
  \bibinfo {author} {\bibfnamefont {D.~R.}\ \bibnamefont {Reichman}}, \ and\
  \bibinfo {author} {\bibfnamefont {J.~D.}\ \bibnamefont {Eaves}},\ }\bibinfo{title}{Theory of coherent resonance energy transfer}, \href
  {\doibase 10.1063/1.2977974} {\bibfield  {journal} {\bibinfo  {journal} {J.
  Chem. Phys.}\ }\textbf {\bibinfo {volume} {129}},\ \bibinfo {eid} {101104}
  (\bibinfo {year} {2008})}\BibitemShut {NoStop}%
\bibitem [{\citenamefont {McCutcheon}\ \emph {et~al.}(2011)\citenamefont
  {McCutcheon}, \citenamefont {Dattani}, \citenamefont {Gauger}, \citenamefont
  {Lovett},\ and\ \citenamefont {Nazir}}]{nazir_prb_2011}%
  \BibitemOpen
  \bibfield  {author} {\bibinfo {author} {\bibfnamefont {D.~P.~S.}\
  \bibnamefont {McCutcheon}}, \bibinfo {author} {\bibfnamefont {N.~S.}\
  \bibnamefont {Dattani}}, \bibinfo {author} {\bibfnamefont {E.~M.}\
  \bibnamefont {Gauger}}, \bibinfo {author} {\bibfnamefont {B.~W.}\
  \bibnamefont {Lovett}}, \ and\ \bibinfo {author} {\bibfnamefont
  {A.}~\bibnamefont {Nazir}},\ }\bibinfo{title}{A general approach to quantum dynamics using a variational master equation: Application to phonon-damped Rabi rotations in quantum dots}, \href {\doibase 10.1103/PhysRevB.84.081305}
  {\bibfield  {journal} {\bibinfo  {journal} {Phys. Rev. B}\ }\textbf {\bibinfo
  {volume} {84}},\ \bibinfo {pages} {081305} (\bibinfo {year}
  {2011})}\BibitemShut {NoStop}%
\bibitem [{\citenamefont {Zimanyi}\ and\ \citenamefont
  {Silbey}(2012)}]{silbey_vari._pola._transf._ptrsa_2012}%
  \BibitemOpen
  \bibfield  {author} {\bibinfo {author} {\bibfnamefont {E.~N.}\ \bibnamefont
  {Zimanyi}}\ and\ \bibinfo {author} {\bibfnamefont {R.~J.}\ \bibnamefont
  {Silbey}},\ }\bibinfo{title}{Theoretical description of quantum effects in multi-chromophoric aggregates}, \href {\doibase 10.1098/rsta.2011.0204} {\bibfield  {journal}
  {\bibinfo  {journal} {Phil. Trans. R. Soc. A}\ }\textbf {\bibinfo {volume}
  {370}},\ \bibinfo {pages} {3620} (\bibinfo {year} {2012})}\BibitemShut
  {NoStop}%
\bibitem [{\citenamefont {Apollaro}\ \emph {et~al.}(2014)\citenamefont
  {Apollaro}, \citenamefont {Lorenzo}, \citenamefont {Di~Franco}, \citenamefont
  {Plastina},\ and\ \citenamefont
  {Paternostro}}]{apollaro_comparison_pra_2014}%
  \BibitemOpen
  \bibfield  {author} {\bibinfo {author} {\bibfnamefont {T.~J.~G.}\
  \bibnamefont {Apollaro}}, \bibinfo {author} {\bibfnamefont {S.}~\bibnamefont
  {Lorenzo}}, \bibinfo {author} {\bibfnamefont {C.}~\bibnamefont {Di~Franco}},
  \bibinfo {author} {\bibfnamefont {F.}~\bibnamefont {Plastina}}, \ and\
  \bibinfo {author} {\bibfnamefont {M.}~\bibnamefont {Paternostro}},\ }\bibinfo{title}{Competition between memory-keeping and memory-erasing decoherence channels}, \href
  {\doibase 10.1103/PhysRevA.90.012310} {\bibfield  {journal} {\bibinfo
  {journal} {Phys. Rev. A}\ }\textbf {\bibinfo {volume} {90}},\ \bibinfo
  {pages} {012310} (\bibinfo {year} {2014})}\BibitemShut {NoStop}%
\bibitem [{\citenamefont {Chen}\ \emph
  {et~al.}(2015{\natexlab{b}})\citenamefont {Chen}, \citenamefont {Lien},
  \citenamefont {Chen},\ and\ \citenamefont
  {Chen}}]{hongbin_k_div_diag_pra_2015}%
  \BibitemOpen
  \bibfield  {author} {\bibinfo {author} {\bibfnamefont {H.-B.}\ \bibnamefont
  {Chen}}, \bibinfo {author} {\bibfnamefont {J.-Y.}\ \bibnamefont {Lien}},
  \bibinfo {author} {\bibfnamefont {G.-Y.}\ \bibnamefont {Chen}}, \ and\
  \bibinfo {author} {\bibfnamefont {Y.-N.}\ \bibnamefont {Chen}},\ }\bibinfo{title}{Hierarchy of non-Markovianity and $k$-divisibility phase diagram of quantum processes in open systems}, \href
  {\doibase 10.1103/PhysRevA.92.042105} {\bibfield  {journal} {\bibinfo
  {journal} {Phys. Rev. A}\ }\textbf {\bibinfo {volume} {92}},\ \bibinfo
  {pages} {042105} (\bibinfo {year} {2015}{\natexlab{b}})}\BibitemShut
  {NoStop}%
\bibitem [{\citenamefont {Scala}\ \emph
  {et~al.}(2007{\natexlab{a}})\citenamefont {Scala}, \citenamefont {Militello},
  \citenamefont {Messina}, \citenamefont {Piilo},\ and\ \citenamefont
  {Maniscalco}}]{piilo_jc_justification_pra_2007}%
  \BibitemOpen
  \bibfield  {author} {\bibinfo {author} {\bibfnamefont {M.}~\bibnamefont
  {Scala}}, \bibinfo {author} {\bibfnamefont {B.}~\bibnamefont {Militello}},
  \bibinfo {author} {\bibfnamefont {A.}~\bibnamefont {Messina}}, \bibinfo
  {author} {\bibfnamefont {J.}~\bibnamefont {Piilo}}, \ and\ \bibinfo {author}
  {\bibfnamefont {S.}~\bibnamefont {Maniscalco}},\ }\bibinfo{title}{Microscopic derivation of the Jaynes-Cummings model with cavity losses}, \href {\doibase
  10.1103/PhysRevA.75.013811} {\bibfield  {journal} {\bibinfo  {journal} {Phys.
  Rev. A}\ }\textbf {\bibinfo {volume} {75}},\ \bibinfo {pages} {013811}
  (\bibinfo {year} {2007}{\natexlab{a}})}\BibitemShut {NoStop}%
\bibitem [{\citenamefont {Scala}\ \emph
  {et~al.}(2007{\natexlab{b}})\citenamefont {Scala}, \citenamefont {Militello},
  \citenamefont {Messina}, \citenamefont {Maniscalco}, \citenamefont {Piilo},\
  and\ \citenamefont {Suominen}}]{piilo_jc_justification_jpa_2007}%
  \BibitemOpen
  \bibfield  {author} {\bibinfo {author} {\bibfnamefont {M.}~\bibnamefont
  {Scala}}, \bibinfo {author} {\bibfnamefont {B.}~\bibnamefont {Militello}},
  \bibinfo {author} {\bibfnamefont {A.}~\bibnamefont {Messina}}, \bibinfo
  {author} {\bibfnamefont {S.}~\bibnamefont {Maniscalco}}, \bibinfo {author}
  {\bibfnamefont {J.}~\bibnamefont {Piilo}}, \ and\ \bibinfo {author}
  {\bibfnamefont {K.-A.}\ \bibnamefont {Suominen}},\ }\bibinfo{title}{Cavity losses for the dissipative Jaynes-Cummings Hamiltonian beyond rotating wave approximation}, \href
  {http://stacks.iop.org/1751-8121/40/i=48/a=015} {\bibfield  {journal}
  {\bibinfo  {journal} {J. Phys. A: Math. Theor.}\ }\textbf {\bibinfo {volume}
  {40}},\ \bibinfo {pages} {14527} (\bibinfo {year}
  {2007}{\natexlab{b}})}\BibitemShut {NoStop}%
\end{thebibliography}

%

\end{document}